\documentclass[12pt]{article}
\usepackage{graphicx}
\usepackage{scicite}
\usepackage{bm}
\usepackage{times}
\usepackage{amsmath}
\usepackage{amssymb}
\usepackage{times}
\usepackage{physics}
\usepackage{hyperref}
\usepackage{relsize}
\usepackage[normalem]{ulem}
\usepackage{xcolor}
\usepackage{totcount}
\regtotcounter{figure}
\regtotcounter{enumiv}
\newenvironment{sciabstract}{%
\begin{quote} \bf}
{\end{quote}}

\begin{document} 
\baselineskip24pt

\title{A $B_{1g}$ phonon unlocks $d$-wave superconductivity in the repulsive Hubbard model}
\author
{Wen O. Wang,$^{1,\ast}$  Tianshu Li,$^{2}$  Xiaoyue Ma,$^{2,3}$  Rong Zhang,$^{3,4}$\\
Lei Zhang,$^{3}$  Shruti Agarwal,$^{5,6}$  Benjamin~Cohen-Stead,$^{5,6}$\\
Emily Z. Zhang,$^{2,7}$  Brian Moritz,$^{3}$\\
Steven Johnston,$^{5,6}$ Zhi-Xun Shen,$^{3,4,7,8,9\ddagger}$  Thomas P. Devereaux,$^{2,3,7,\dagger}$\\
\\
\normalsize{$^{1}$Kavli Institute for Theoretical Physics, University of California,}\\
\normalsize{Santa Barbara, CA 93106, USA}\\
\normalsize{$^{2}$Department of Materials Science and Engineering, Stanford University,}\\
\normalsize{Stanford, CA 94305, USA}\\
\normalsize{$^{3}$Stanford Institute for Materials and Energy Sciences,}\\
\normalsize{SLAC National Accelerator Laboratory, 2575 Sand Hill Road,}\\
\normalsize{Menlo Park, CA 94025, USA}\\
\normalsize{$^{4}$Department of Applied Physics, Stanford University, Stanford, CA 94305, USA} \\
\normalsize{$^{5}$Department of Physics and Astronomy, The University of Tennessee,}\\
\normalsize{Knoxville, TN 37996, USA}\\
\normalsize{$^{6}$Institute for Advanced Materials and Manufacturing,}\\
\normalsize{The University of Tennessee, Knoxville, TN 37996, USA}\\
\normalsize{$^{7}$Geballe Laboratory for Advanced Materials, Stanford University,}\\
\normalsize{Stanford, CA 94305, USA}\\
\normalsize{$^{8}$Department of Physics, Stanford University, Stanford, CA 94305, USA} \\
\normalsize{$^{9}$Stanford Laboratory for Intelligent Matter, Stanford University, Stanford, CA 94305, USA} \\
\normalsize{$^\ast$E-mail: wenwang.physics@gmail.com.}\\
\normalsize{$^\ddagger$E-mail: zxshen@stanford.edu.}\\
\normalsize{$^\dagger$E-mail: tpd@stanford.edu.}
}

\maketitle

\clearpage
\begin{sciabstract}
\emergencystretch=1em
Using numerically exact finite-temperature quantum Monte Carlo, we demonstrate that a $\boldsymbol{B_{1g}}$ phonon coupled to the doped $\boldsymbol{t}$--$\boldsymbol{t'}$ repulsive Hubbard model produces genuine $\boldsymbol{d}$-wave superconductivity at a substantial transition temperature, in excess of either interaction alone.
The observed trends suggest that stronger Hubbard repulsion may cooperate with the phonon to further raise the pairing scale.
The enhancement persists at stronger repulsion within the accessible window, limited by the fermion sign problem, and follows a common trend across phonon frequencies $\boldsymbol{\Omega}$ when organized by a retardation scale $\boldsymbol{2\pi T/\Omega}$.
We also find a well-defined superconducting dome in doping, echoing a characteristic feature of correlated superconductors.
Our results identify the $\boldsymbol{B_{1g}}$ phonon as a plausible microscopic ingredient for robust $\boldsymbol{d}$-wave superconductivity in the Hubbard model and suggest a concrete, physically motivated route toward materials with higher superconducting transition temperatures.
\end{sciabstract}

High-temperature superconductivity has long fascinated researchers within and beyond con\-densed-matter physics, both for its rich quantum many-body physics and for its far-reaching technological potential. An established experimental fact in the cuprates is the $d$-wave pairing symmetry~\cite{PhysRevLett.70.1553,PhysRevLett.70.3999,PhysRevLett.71.2134,PhysRevLett.73.593}, suggesting that strong Coulomb interactions are an important ingredient.
However, despite decades of effort, we still lack a settled microscopic picture of why these materials superconduct at such high temperatures and which microscopic ingredients ultimately control their pairing scale.
For much of this history, the repulsive Hubbard interaction has stood at the center of the discussion~\cite{Hubbard1963,Anderson1987,annurev:/content/journals/10.1146/annurev-conmatphys-031620-102024,annurev:/content/journals/10.1146/annurev-conmatphys-090921-033948}.
The possibility that repulsion alone can organize electrons into coherent $d$-wave pairs, and produce superconductivity with a sizable transition temperature, is conceptually remarkable and has driven decades of effort to establish superconductivity in the Hubbard model~\cite{RevModPhys.84.1383}.
Yet, robust $d$-wave superconductivity in the pure Hubbard model has proved surprisingly difficult to establish~\cite{PhysRevB.109.085121,jiang2025competitionchargedensitywavesuperconductingorders}.
Controlled calculations reveal a landscape crowded with competing tendencies, particularly intertwined charge and spin stripes~\cite{PhysRevX.11.031007,Huang2018,doi:10.1073/pnas.2112806119,PhysRevB.107.085126,doi:10.1126/science.aal5304,PhysRevB.109.085121,jiang2025competitionchargedensitywavesuperconductingorders}.
Finite-temperature exact quantum Monte Carlo finds pronounced stripe correlations, particularly on the hole-doped side~\cite{Huang2018,PhysRevB.107.085126,Wang2025FiniteTemperature}, while ground-state density matrix renormalization group (DMRG) shows strong sensitivity to geometry and band parameters, with competing states separated by energy differences as small as $10^{-3}t$~\cite{doi:10.1126/science.aam7127,doi:10.1126/science.aal5304,PhysRevB.109.085121,jiang2025competitionchargedensitywavesuperconductingorders}.
Increasingly sophisticated numerical work points toward the possibility that the Hubbard model may be missing essential ingredients for robust superconductivity.
Especially given the near-degeneracy of these competing states, even a relatively weak additional interaction may tip the balance in favor of superconductivity.

Phonons provide the conventional mechanism for superconductivity~\cite{PhysRev.108.1175}, yet historically they have occupied a relatively marginal place in many theories of cuprate superconductivity, in part because the strongly correlated parent state and $d_{x^2-y^2}$ pairing symmetry have motivated a focus on electronic pairing mechanisms.
Although electron--phonon interactions generally favor $s$-wave pairing, certain phonon modes can also provide an attractive projection in the $d$-wave channel~\cite{AndersenSavrasovJepsenLiechtenstein1996,JepsenAndersenDasguptaSavrasov1998,PhysRevB.75.014503}.
In this work, we focus on the $B_{1g}$ oxygen buckling mode (Fig.~\ref*{fig:main_pairing}A), an out-of-plane vibration of the planar oxygen, which may provide a particularly effective route to $d$-wave superconductivity.
Signatures of the $B_{1g}$ oxygen buckling mode have been reported in a variety of experiments on the cuprates.
In particular, Raman measurements across doping indicate substantial coupling of this mode to the electronic degrees of freedom~\cite{PhysRevB.60.9836}, while ARPES and microscopic calculations show that this coupling is strongly weighted toward the antinodal region, where the $d$-wave superconducting gap is maximal~\cite{PhysRevB.51.505,PhysRevLett.93.117003,PhysRevLett.93.117004}.
These findings motivate including the $B_{1g}$ phonon as a realistic microscopic ingredient in modeling cuprate superconductivity.

Within the Zhang--Rice-singlet description~\cite{PhysRevB.37.3759}, frozen-phonon exact diagonalization of multiorbital Cu--O clusters found a substantial first-order modulation of the effective hopping by the $B_{1g}$ distortion, while the local energy changes only at second order~\cite{PhysRevB.82.064513}.
Motivated by these findings, we study a $t$--$t'$--$U$ Hubbard model coupled to a $B_{1g}$ phonon through hopping modulation,
\begin{equation}
\begin{aligned}
H&=H_{\rm Hub}+H_{\rm ph}-g\sum_i Q_i\mathcal K_{B_{1g},i},\\
\mathcal K_{B_{1g},i}&=\tfrac12\bigl(K_{i,i+\hat x}+K_{i,i-\hat x}-K_{i,i+\hat y}-K_{i,i-\hat y}\bigr).
\end{aligned}
\label{eq:b1g_model}
\end{equation}
Here $K_{ij}=\sum_\sigma(c_{i\sigma}^\dagger c_{j\sigma}+\mathrm{H.c.})$ is the bond-hopping operator and $Q_i$ is the local buckling coordinate, with $g$ the coupling strength.
Each $Q_i$ modulates the four bonds surrounding site $i$, in contrast to the independent bond phonons in the Su--Schrieffer--Heeger--Hubbard model, in which DMRG found robust $d$-wave superconductivity~\cite{WangJiangYao2025}, further motivating us to examine how hopping modulation affects pairing.
In our model, the opposite signs along $x$ and $y$ give the electron--phonon coupling $B_{1g}$ symmetry, with a $d_{x^2-y^2}$ form factor proportional to $\cos k_x-\cos k_y$ for a uniform ($\mathbf q=0$) distortion (see Eq.~\eqref{eq:b1g_momentum_vertex} in the Supplementary Materials).
In contrast, a nearest-neighbor density coupling, allowed by symmetry alone, has a vertex proportional to $\cos q_x-\cos q_y$ and vanishes at the zone center. It therefore does not reproduce the long-wavelength $B_{1g}$ coupling derived in Ref.~\cite{PhysRevB.82.064513}.
$H_{\rm Hub}=-t\sum_{\langle ij\rangle}K_{ij}-t'\sum_{\langle\!\langle ij\rangle\!\rangle}K_{ij}+U\sum_i n_{i\uparrow}n_{i\downarrow}-\mu\sum_i n_i$ is the $t$--$t'$--$U$ Hubbard Hamiltonian, where $t$ ($t'=-0.25t$) is the nearest-neighbor (next-nearest-neighbor) hopping energy, $U$ is the on-site interaction, and $\mu$ is the chemical potential controlling the electron density, with our main focus on $\langle n\rangle=0.8$; $H_{\rm ph}=\sum_i[P_i^2/(2M)+M\Omega^2Q_i^2/2]$ is the harmonic phonon Hamiltonian, where $M$ and $\Omega$ are the phonon mass and frequency and $P_i$ is conjugate to $Q_i$.
We define the coupling parameter as $\lambda=g^2/(2Mt\Omega^2)$ and use $\lambda=0.2$~\cite{PhysRevLett.93.117004,PhysRevB.82.064513} for all calculations with phonons.

Here, using determinant quantum Monte Carlo (DQMC), we report that a $B_{1g}$ phonon and on-site Hubbard repulsion jointly produce robust $d$-wave superconductivity at a sizable transition temperature, with both ingredients essential in the regime studied.
Despite the fermion sign problem, the transition is accessible to finite-temperature quantum Monte Carlo at weak repulsion: we find a transition temperature scale of about $0.04t$ at $U/t=1$ and $\langle n\rangle=0.8$, with higher scales across the doping dome at larger $U$.
At even larger $U$, the increasingly severe sign problem restricts low temperature access~\cite{PhysRevB.41.9301}, preventing an equally direct demonstration of superconductivity.
Nevertheless, the enhancement of the pairing vertex depends only weakly on $U$, with hints of an increase with repulsion at low temperatures.
These findings suggest that the phonon and on-site repulsion may cooperate beyond simple additivity to enhance the pairing energy scale and further raise the transition temperature.
Together, our results identify the $B_{1g}$ phonon as a realistic missing ingredient for robust $d$-wave superconductivity in the repulsive Hubbard model and offer a concrete direction for engineering higher superconducting scales.

\section*{A $\boldsymbol{B_{1g}}$ phonon and Hubbard repulsion\\unlock $\boldsymbol{d}$-wave superconductivity}

The $B_{1g}$ phonon promotes pairing in both a $d$-wave channel with form factor $f_d(\mathbf k)=\cos k_x-\cos k_y$ and an $A_{1g}$ sector containing on-site and extended $s$-wave components, spanned by $1$ and $f_d^2(\mathbf k)$ (see Eq.~\eqref{eq:b1g_pair_channels} in the Supplementary Materials), but on-site Hubbard repulsion $U$ suppresses the competing $A_{1g}$ components.
For the parameter regimes studied here, the $d$-wave channel already dominates over the competing $A_{1g}$ channels for $U/t\gtrsim0.1$, as shown in Supplementary Figs.~\ref{fig:supp_a1g} and~\ref{fig:supp_uc_vertex}.
Fig.~\ref{fig:main_pairing}B--D therefore focuses on $U/t=1$, where the fermion sign is manageable so that the $d$-wave superconducting regime is accessible, with the phonon frequency fixed at $\Omega/t=4$.

We establish superconductivity using two complementary observables.
The first is the imaginary-time midpoint pairing susceptibility $\chi_d=(\beta/N_s)\langle\mathcal T_\tau\Delta_d(\beta/2)\Delta_d^\dagger(0)\rangle$, constructed from the pairing correlation at $\tau=\beta/2$ to emphasize the low-energy pairing response~\cite{Wang2025FiniteTemperature,chen2025topological} (see also Supplementary Fig.~\ref{fig:supp_estimator}).
Here $\beta=1/T$ is the inverse temperature, $N_s$ is the number of sites, and $\Delta_d$ is the $d$-wave pair annihilation operator.
We focus on the temperature dependence of $\chi_d$ in Fig.~\ref{fig:main_pairing}B over the temperature range where finite-size effects are small, and on its finite-size scaling closer to the transition in Fig.~\ref{fig:main_pairing}D.
The second is the superfluid density $\rho_s$~\cite{PhysRevLett.68.2830} in Fig.~\ref{fig:main_pairing}C (see the Formalism section in the Supplementary Materials).

In Fig.~\ref{fig:main_pairing}B, $\chi_d^{-1}$ decreases rapidly toward zero upon cooling.
Using only data from the temperature range where the $8\times8$ and $10\times10$ results nearly coincide and thus approximate the thermodynamic-limit response, we extrapolate the $10\times10$ inverse susceptibility to zero at the Curie--Weiss temperature $T_0/t\approx0.04$.
Although obtained at much weaker repulsion, this transition scale is an order of magnitude larger than the small energy-density differences reported between competing ordered states in the pure Hubbard model at stronger repulsion~\cite{doi:10.1126/science.aam7127,doi:10.1126/science.aal5304,PhysRevB.109.085121,jiang2025competitionchargedensitywavesuperconductingorders}.
By contrast, neither the pure Hubbard model ($U/t=1$, $\lambda=0$) nor the phonon-coupled model without Hubbard repulsion ($U=0$, $\lambda=0.2$) exhibits this signature of a $d$-wave superconducting transition.
This contrast shows that the $d$-wave superconductivity observed here requires the joint action of the $B_{1g}$ phonon and Hubbard repulsion.

Both the superfluid density $\rho_s$ in Fig.~\ref{fig:main_pairing}C and the finite-size scaling of $\chi_d$ in Fig.~\ref{fig:main_pairing}D support a transition near $T_0$.
In two dimensions, the Berezinskii--Kosterlitz--Thouless (BKT) transition is characterized by the universal relation $\rho_s(T_c^-)=2T_c/\pi$ in the thermodynamic limit~\cite{KosterlitzThouless1973,NelsonKosterlitz1977}.
On the $6\times6$ lattice, $\rho_s$ rises upon cooling and closely approaches the BKT reference line $2T/\pi$ near $T_0$.
The fermion sign problem prevents the $8\times8$ and $10\times10$ lattices from reaching equally low temperatures, but their sharper low-temperature upturns are consistent with crossings near the same temperature.
At $T_c$, BKT theory also predicts the leading finite-size scaling $\chi_d\sim L^{7/4}$.
In Fig.~\ref{fig:main_pairing}D, the rescaled response $T\chi_d/L^{7/4}$ for the $6\times6$ and $8\times8$ lattices already approaches a common value above $T_0$, while the $10\times10$ trend suggests a crossing at a lower temperature, closer to the estimate from Fig.~\ref{fig:main_pairing}B.
Together, these three criteria identify a $d$-wave superconducting transition, with a consistent estimate of $T_c/t\approx0.04$.
The suppression of the low-temperature spin response, together with a finite charge compressibility, further supports singlet pairing (see Supplementary Fig.~\ref{fig:supp_thermodynamics}).
To our knowledge, this is the first numerically exact DQMC calculation to directly resolve a finite-temperature $d$-wave superconducting transition in a repulsive Hubbard system coupled to dynamical phonons, without an explicitly added pairing interaction.

\section*{Cooperative enhancement of $\boldsymbol{d}$-wave pairing}

The sizable $T_c$ produced by combining the $B_{1g}$ phonon and Hubbard repulsion is striking: conventional weak-coupling phonon mechanisms yield small transition scales~\cite{PhysRev.108.1175,PhysRevB.82.064513}, while $d$-wave superconductivity in the pure Hubbard model faces strong competition from other orders~\cite{PhysRevB.109.085121,PhysRevX.11.031007}.
How do these two ingredients cooperate to produce this high transition scale?
To address this question, we examine how the pairing scale evolves with $U$ and what this implies for $T_c$.
In particular, we ask whether $d$-wave superconductivity persists as the Hubbard repulsion $U$ increases toward the stronger-coupling regime more relevant to the cuprates.

Figure~\ref{fig:main_vertex}A addresses this evolution through the dimensionless vertex enhancement ratio $R_d=\bar{\chi}_d/\chi_d-1$, for which $R_d<0$ signifies an attractive $d$-wave vertex~\cite{PhysRevB.39.839,PhysRevB.90.075121}.
At high temperatures, $R_d$ lies near zero, where the pair response is dominated by the single-particle bubble $\bar\chi_d$; the phonon-coupled curves become increasingly negative upon cooling, suggesting strong $d$-wave enhancement.
Remarkably, the phonon enhancement persists across the full interaction range studied: at all $U$, the $B_{1g}$ phonon drives $R_d$ substantially more negative than in the corresponding pure Hubbard model, producing a much stronger $d$-wave pairing tendency.
The phonon also produces a more extended $d$-wave pattern in the real-space pair correlations at both weak and stronger $U$ (see Supplementary Fig.~\ref{fig:supp_realspace}).
Despite the progressively restricted temperature range imposed by the fermion sign problem at larger $U$, the magnitude of the attraction shows weak dependence on $U$ up to around $U/t=6$ at the lowest common temperature accessible across the wide range of $U$ studied.

In fact, the low-temperature comparison between $U/t=1$ and $2$ already points to a higher transition temperature scale at $U/t=2$.
For still larger $U$, $R_d$ is progressively less negative at high temperatures, but the curves draw closer to those at weaker repulsion upon cooling within the increasingly restricted temperature window.
This evolution hints that the larger-$U$ curves may cross below the weaker-$U$ curves at lower temperatures and approach $R_d=-1$ at a higher transition scale.
Together, these trends suggest that the low-temperature pairing enhancement seen at $U/t=1$ and $2$ not only persists to larger $U$, but is also accompanied by a further increase in the pairing scale over a substantial range of $U$.

This behavior goes beyond the role of $U$ in the bare interaction analysis. The on-site repulsion suppresses the competing $A_{1g}$ channels but has no direct projection onto the $d_{x^2-y^2}$ form factor, leaving the bare $d$-wave attraction independent of $U$ (see Eq.~\eqref{eq:b1g_pair_channels} in the Supplementary Materials).
Here, however, the phonon and electronic correlations cooperate in strengthening pairing.
To quantify the phonon's ``net enhancement'', Fig.~\ref{fig:main_vertex}B plots \mbox{$\chi_d(\lambda=0.2)/\chi_d(\lambda=0)$}, which factors out the pure Hubbard pairing response at the same $U$ and temperature.
Upon cooling, the enhancement increases sharply from $U=0$ to $U/t=1$, where the competing $A_{1g}$ channels are suppressed in favor of $d$-wave pairing.
At comparable temperatures, the enhancement increases further with $U$.
Thus, even at the same bare phonon coupling, the phonon enhancement of $d$-wave pairing grows with $U$.
An analysis of the effective $d$-wave pairing interaction further shows that, at low temperatures, the combined attraction exceeds the sum of the separate phonon and Hubbard contributions (see Supplementary Fig.~\ref{fig:supp_nonadditivity}).
Taken together, these findings demonstrate that the cooperation between the phonon and Hubbard repulsion goes beyond simple additivity, suggesting a possible route to a higher $T_c$.

\section*{Retardation sets the pairing scale}

We now examine how the phonon frequency affects superconductivity.
As shown in Fig.~\ref{fig:main_omega_crossover}A and Supplementary Fig.~\ref{fig:supp_vertex_omega}, the pairing vertex exhibits a pronounced dependence on the phonon frequency $\Omega$.
At fixed $\lambda=0.2$, $U$, and filling, increasing $\Omega$ makes the $d$-wave vertex more attractive at a given temperature, with $R_d$ becoming more negative, suggesting a higher transition scale.

To identify the temperature scale at which the phonon enhancement sets in, we plot the enhancement ratio $\chi_d(\lambda=0.2)/\chi_d(\lambda=0)$ versus $2\pi T/\Omega$ in Fig.~\ref{fig:main_omega_crossover}B--D.
Notably, the enhancement organizes around a common retardation scale across both $U$ and $\Omega$.
The phonon enhancement, manifested as $\chi_d(\lambda=0.2)/\chi_d(\lambda=0)\gtrsim1$, appears only when the lowest nonzero bosonic Matsubara frequency, $2\pi T$, falls below the phonon frequency $\Omega$ (see also Supplementary Fig.~\ref{fig:supp_omega_crossover}).
Above this scale, scattering from the increasingly quasi-static phonon can reduce low-energy single-particle weight and thereby suppress the full $d$-wave pair response.
Once $2\pi T\lesssim\Omega$, finite-frequency phonon-mediated attraction from the retarded interaction (Eq.~\eqref{eq:b1g_retarded_interaction} in the Supplementary Materials) becomes effective and the pairing enhancement grows rapidly upon cooling.
At small $\Omega$ and weak $U$, the ratios in Fig.~\ref{fig:main_omega_crossover}B and Supplementary Fig.~\ref{fig:supp_omega_crossover}A deviate downward from this common trend. At these frequencies and temperatures, the ratios show substantial finite-size effects that may account for the deviation (Supplementary Fig.~\ref{fig:supp_size_ratio}).

The similar scaling of the phonon enhancement across $U$ and $\Omega$ offers two broader insights. First, previous studies, including calculations of the cuprate $B_{1g}$ phonon, have reached differing conclusions about whether electron--phonon coupling enhances or suppresses $d$-wave pairing~\cite{PhysRevB.75.014503,Macridin2012JPCM}. Our results identify retardation as an important part of this apparent discrepancy: for slow phonons, temperatures must fall below the phonon scale, $2\pi T\lesssim\Omega$, before their pairing enhancement becomes apparent. Second, experimentally relevant regimes may combine relatively low phonon frequencies (which require correspondingly lower temperatures to reveal their pairing enhancement) with stronger electronic correlations, precisely where unbiased finite-temperature simulations become most difficult. Yet throughout the broad range of $U$ and $\Omega$ accessible here, the enhancement behavior follows a common pattern across the parameter space. This robustness across both interaction strength and phonon frequency suggests that our results may generalize to regimes relevant to real materials.

\section*{The superconducting dome}

Extending the inverse-susceptibility extrapolation introduced in Fig.~\ref{fig:main_pairing}B to a wider doping range, we determine $T_0$ for $U/t=1$ and $2$, focusing on $\Omega/t=4$, as shown in Fig.~\ref{fig:main_doping}A (see Supplementary Fig.~\ref{fig:supp_u_tc}A,B for extrapolation details).
We directly observe the trend toward a higher transition scale with increasing Hubbard repulsion discussed above: $T_0$ is higher at $U/t=2$ than at $U/t=1$ across most of the doping range studied, especially near optimal doping, $1-\langle n\rangle\approx15$--$20\%$.
Figure~\ref{fig:main_doping}B uses $10\times10$ data to extrapolate $T_0$ and focuses on the $U$ dependence at fixed $\langle n\rangle=0.8$ (see Supplementary Fig.~\ref{fig:supp_u_tc}C,D). We find an increase in $T_0$ with $U$, and in particular, using the same fitting criterion, we find that the phonon-only model at $U=0$ has a much lower $T_0$ in its leading $A_{1g}$ channel than the $d$-wave values at $U/t=1$ and $2$. This comparison shows quantitatively that, in this regime, a sizable transition scale requires cooperation between the phonon and Hubbard repulsion.

\begingroup
\emergencystretch=1em
Remarkably, we also observe a well-defined superconducting dome at both interaction strengths, $U/t=1$ and $2$, with $T_0$ decreasing on either side of optimal doping.
The pronounced doping dependence is consistent with superconducting domes observed experimentally in many correlated superconductors~\cite{RevModPhys.84.1383,RevModPhys.83.1589} and is striking at such weak Hubbard repulsions, below the strong-coupling Mott regime of the pure Hubbard model.
This doping dependence points to an interplay among the $B_{1g}$ phonon, electronic correlations, and the underlying electronic structure in setting the superconducting scale.
\par
\endgroup

\section*{Cooperative pairing: summary and perspectives}

In this work, we discover that a $B_{1g}$ phonon and on-site Hubbard repulsion jointly produce genuine finite-temperature $d$-wave superconductivity in the doped Hubbard model.
At weak repulsion and large phonon frequencies, the sizable transition temperature brings the superconducting regime within reach of numerically exact quantum Monte Carlo despite the fermion sign problem.
Importantly, our work extends the Hubbard model, in which superconductivity has proved difficult to establish, into a regime of robust $d$-wave superconductivity.
The phonon and Hubbard repulsion cooperate to enhance pairing, pointing toward robust superconductivity at larger $U$.
We also observe a well-defined superconducting dome in doping, echoing a characteristic feature of correlated superconductors~\cite{RevModPhys.84.1383,RevModPhys.83.1589}.
We further identify a common retardation scale, $2\pi T\sim\Omega$, that governs when the phonon enhances pairing, across interaction strengths and phonon frequencies.
Our results identify the $B_{1g}$ phonon as a plausible missing ingredient that can help tip the near-degenerate Hubbard landscape toward robust $d$-wave superconductivity, and offer a possible direction for engineering higher superconducting scales.

An important question is whether these observations are relevant to high-$T_c$ materials, such as the cuprates.
We are studying quite a simplistic model compared with these materials.
Moreover, the parameter regime of interest involves much lower phonon frequencies, typically about $\Omega/t=0.1$--$0.2$, and stronger repulsion, at least $U/t\sim6$--$8$.
Our study does not establish superconductivity in this regime.
In particular, at larger $U$, stripes and other competing orders may challenge the superconducting state~\cite{PhysRevB.109.085121,PhysRevX.11.031007}.
Nevertheless, the observed cooperative enhancement and common retardation behavior give reason for optimism that this behavior may extend more broadly in parameter space, motivating further computational efforts in these regimes.
Ground-state calculations of the phonon-coupled model, such as DMRG, are needed to determine whether the $B_{1g}$ phonon can stabilize $d$-wave superconductivity against these competing orders and to clarify the origin of pairing at strong coupling.

More generally, the interplay between strong electronic correlations and electron--phonon coupling is an important topic for further investigation~\cite{Klupp2012}.
Experimental reports include substantial enhancement of superconductivity near the metal--Mott-insulator boundary in fullerides~\cite{Potocnik2014,Zadik2015} and correlation-enhanced electron--phonon coupling in FeSe~\cite{Gerber2017}.
Our study suggests that this rich area of physics may become accessible to numerical studies.

\begingroup
\baselineskip22pt
\bibliography{main}

\begin{thebibliography}{10}

\bibitem{PhysRevLett.70.1553}
Z.-X. Shen, {\it et~al.\/}, {\it Phys. Rev. Lett.\/} {\bf 70}, 1553 (1993).

\bibitem{PhysRevLett.70.3999}
W.~N. Hardy, D.~A. Bonn, D.~C. Morgan, R.~Liang, K.~Zhang, {\it Phys. Rev. Lett.\/} {\bf 70}, 3999 (1993).

\bibitem{PhysRevLett.71.2134}
D.~A. Wollman, D.~J. Van~Harlingen, W.~C. Lee, D.~M. Ginsberg, A.~J. Leggett, {\it Phys. Rev. Lett.\/} {\bf 71}, 2134 (1993).

\bibitem{PhysRevLett.73.593}
C.~C. Tsuei, {\it et~al.\/}, {\it Phys. Rev. Lett.\/} {\bf 73}, 593 (1994).

\bibitem{Hubbard1963}
J.~Hubbard, {\it Proc. R. Soc. Lond. A\/} {\bf 276}, 238 (1963).

\bibitem{Anderson1987}
P.~W. Anderson, {\it Science\/} {\bf 235}, 1196 (1987).

\bibitem{annurev:/content/journals/10.1146/annurev-conmatphys-031620-102024}
D.~P. Arovas, E.~Berg, S.~A. Kivelson, S.~Raghu, {\it Annual Review of Condensed Matter Physics\/} {\bf 13}, 239 (2022).

\bibitem{annurev:/content/journals/10.1146/annurev-conmatphys-090921-033948}
M.~Qin, T.~Schäfer, S.~Andergassen, P.~Corboz, E.~Gull, {\it Annual Review of Condensed Matter Physics\/} {\bf 13}, 275 (2022).

\bibitem{RevModPhys.84.1383}
D.~J. Scalapino, {\it Rev. Mod. Phys.\/} {\bf 84}, 1383 (2012).

\bibitem{PhysRevB.109.085121}
Y.-F. Jiang, T.~P. Devereaux, H.-C. Jiang, {\it Phys. Rev. B\/} {\bf 109}, 085121 (2024).

\bibitem{jiang2025competitionchargedensitywavesuperconductingorders}
H.-C. Jiang, T.~P. Devereaux, S.~A. Kivelson, {\it npj Quantum Materials\/} {\bf 11}, 80 (2026).

\bibitem{PhysRevX.11.031007}
A.~Wietek, Y.-Y. He, S.~R. White, A.~Georges, E.~M. Stoudenmire, {\it Phys. Rev. X\/} {\bf 11}, 031007 (2021).

\bibitem{Huang2018}
E.~W. Huang, C.~B. Mendl, H.-C. Jiang, B.~Moritz, T.~P. Devereaux, {\it npj Quantum Materials\/} {\bf 3}, 22 (2018).

\bibitem{doi:10.1073/pnas.2112806119}
P.~Mai, S.~Karakuzu, G.~Balduzzi, S.~Johnston, T.~A. Maier, {\it Proceedings of the National Academy of Sciences\/} {\bf 119}, e2112806119 (2022).

\bibitem{PhysRevB.107.085126}
E.~W. Huang, {\it et~al.\/}, {\it Phys. Rev. B\/} {\bf 107}, 085126 (2023).

\bibitem{doi:10.1126/science.aal5304}
H.-C. Jiang, T.~P. Devereaux, {\it Science\/} {\bf 365}, 1424 (2019).

\bibitem{Wang2025FiniteTemperature}
W.~O. Wang, T.~P. Devereaux, {\it arXiv preprint arXiv:2510.16616\/}  (2025).

\bibitem{doi:10.1126/science.aam7127}
B.-X. Zheng, {\it et~al.\/}, {\it Science\/} {\bf 358}, 1155 (2017).

\bibitem{PhysRev.108.1175}
J.~Bardeen, L.~N. Cooper, J.~R. Schrieffer, {\it Phys. Rev.\/} {\bf 108}, 1175 (1957).

\bibitem{AndersenSavrasovJepsenLiechtenstein1996}
O.~K. Andersen, S.~Y. Savrasov, O.~Jepsen, A.~I. Liechtenstein, {\it J. Low Temp. Phys.\/} {\bf 105}, 285 (1996).

\bibitem{JepsenAndersenDasguptaSavrasov1998}
O.~Jepsen, O.~K. Andersen, I.~Dasgupta, S.~Savrasov, {\it J. Phys. Chem. Solids\/} {\bf 59}, 1718 (1998).

\bibitem{PhysRevB.75.014503}
C.~Honerkamp, H.~C. Fu, D.-H. Lee, {\it Phys. Rev. B\/} {\bf 75}, 014503 (2007).

\bibitem{PhysRevB.60.9836}
M.~Opel, {\it et~al.\/}, {\it Phys. Rev. B\/} {\bf 60}, 9836 (1999).

\bibitem{PhysRevB.51.505}
T.~P. Devereaux, A.~Virosztek, A.~Zawadowski, {\it Phys. Rev. B\/} {\bf 51}, 505 (1995).

\bibitem{PhysRevLett.93.117003}
T.~Cuk, {\it et~al.\/}, {\it Phys. Rev. Lett.\/} {\bf 93}, 117003 (2004).

\bibitem{PhysRevLett.93.117004}
T.~P. Devereaux, T.~Cuk, Z.-X. Shen, N.~Nagaosa, {\it Phys. Rev. Lett.\/} {\bf 93}, 117004 (2004).

\bibitem{PhysRevB.37.3759}
F.~C. Zhang, T.~M. Rice, {\it Phys. Rev. B\/} {\bf 37}, 3759 (1988).

\bibitem{PhysRevB.82.064513}
S.~Johnston, {\it et~al.\/}, {\it Phys. Rev. B\/} {\bf 82}, 064513 (2010).

\bibitem{WangJiangYao2025}
H.-X. Wang, Y.-F. Jiang, H.~Yao, {\it Science Bulletin\/} {\bf 70}, 2260 (2025).

\bibitem{PhysRevB.41.9301}
E.~Y. Loh, Jr., {\it et~al.\/}, {\it Phys. Rev. B\/} {\bf 41}, 9301 (1990).

\bibitem{chen2025topological}
F.~Chen, W.~O. Wang, J.-X. Zhang, L.~Balents, D.~N. Sheng, {\it Phys. Rev. Lett.\/} {\bf 136}, 086503 (2026).

\bibitem{PhysRevLett.68.2830}
D.~J. Scalapino, S.~R. White, S.~C. Zhang, {\it Phys. Rev. Lett.\/} {\bf 68}, 2830 (1992).

\bibitem{KosterlitzThouless1973}
J.~M. Kosterlitz, D.~J. Thouless, {\it J. Phys. C: Solid State Phys.\/} {\bf 6}, 1181 (1973).

\bibitem{NelsonKosterlitz1977}
D.~R. Nelson, J.~M. Kosterlitz, {\it Phys. Rev. Lett.\/} {\bf 39}, 1201 (1977).

\bibitem{PhysRevB.39.839}
S.~R. White, D.~J. Scalapino, R.~L. Sugar, N.~E. Bickers, R.~T. Scalettar, {\it Phys. Rev. B\/} {\bf 39}, 839 (1989).

\bibitem{PhysRevB.90.075121}
T.~Ying, {\it et~al.\/}, {\it Phys. Rev. B\/} {\bf 90}, 075121 (2014).

\bibitem{Macridin2012JPCM}
A.~Macridin, B.~Moritz, M.~Jarrell, T.~Maier, {\it J. Phys.: Condens. Matter\/} {\bf 24}, 475603 (2012).

\bibitem{RevModPhys.83.1589}
G.~R. Stewart, {\it Rev. Mod. Phys.\/} {\bf 83}, 1589 (2011).

\bibitem{Klupp2012}
G.~Klupp, {\it et~al.\/}, {\it Nature Communications\/} {\bf 3}, 912 (2012).

\bibitem{Potocnik2014}
A.~Poto{\v{c}}nik, {\it et~al.\/}, {\it Scientific Reports\/} {\bf 4}, 4265 (2014).

\bibitem{Zadik2015}
R.~H. Zadik, {\it et~al.\/}, {\it Science Advances\/} {\bf 1}, e1500059 (2015).

\bibitem{Gerber2017}
S.~Gerber, {\it et~al.\/}, {\it Science\/} {\bf 357}, 71 (2017).

\bibitem{DQMC1}
R.~Blankenbecler, D.~J. Scalapino, R.~L. Sugar, {\it Phys. Rev. D\/} {\bf 24}, 2278 (1981).

\bibitem{Hirsch1983HS}
J.~E. Hirsch, {\it Phys. Rev. B\/} {\bf 28}, 4059 (1983).

\bibitem{FeynmanHibbs1965}
R.~P. Feynman, A.~R. Hibbs, {\it Quantum Mechanics and Path Integrals\/} (McGraw-Hill, New York, 1965).

\bibitem{Johnston2013HubbardHolstein}
S.~Johnston, {\it et~al.\/}, {\it Phys. Rev. B\/} {\bf 87}, 235133 (2013).

\bibitem{SmoQyDQMC1}
B.~Cohen-Stead, {\it et~al.\/}, {\it SciPost Phys. Codebases\/} {\bf 29-v2.0} (2026).

\bibitem{SmoQyDQMC2}
B.~Cohen-Stead, {\it et~al.\/}, {\it SciPost Phys. Codebases\/} {\bf 29-r2.0} (2026).

\bibitem{PhysRev.110.769}
K.~Yosida, {\it Phys. Rev.\/} {\bf 110}, 769 (1958).

\bibitem{PhysRevB.47.14599}
N.~Bulut, D.~J. Scalapino, S.~R. White, {\it Phys. Rev. B\/} {\bf 47}, 14599 (1993).

\bibitem{Bormann1991FiniteSize}
D.~Bormann, T.~Schneider, M.~Frick, {\it Europhys. Lett.\/} {\bf 14}, 101 (1991).

\end{thebibliography}
\endgroup
\bibliographystyle{Science}

\clearpage
\section*{Acknowledgements}
\begin{sloppypar}
We acknowledge helpful discussions with H.-C.~Jiang, S.~Jiang, S.~A.~Kivelson, D.-H.~Lee, N.~Nagaosa, D.~J.~Scalapino, and Y.~Wang.
W.O.W. also acknowledges helpful discussions with J.-X.~Zhang.
\end{sloppypar}

The work at Stanford University and SLAC National Accelerator Laboratory was supported by the U.S. Department of Energy, Office of Basic Energy Sciences, Division of Materials Sciences and Engineering.
W.O.W. acknowledges support from the Simons Collaboration on Ultra-Quantum Matter, which is a grant from the Simons Foundation (Grant No. 651440).
W.O.W. also acknowledges support from NSF Grant No. PHY-2309135 to the KITP and from the Mitchel Postdoctoral Scholar Career Development Fund. E.Z.Z. acknowledges support from the Geballe Laboratory for Advanced Materials (GLAM) Postdoctoral Fellowship. 
Work at the University of Tennessee was supported by the National Science Foundation under Grant No. DMR-2401388.
Computational work was performed on the Sherlock cluster at Stanford University and on resources of the National Energy Research Scientific Computing Center, a U.S. Department of Energy, Office of Science User Facility, using NERSC award BES-ERCAP0031425.

\section*{Code and data availability}
We cross-checked our results using independent DQMC implementations.
The processed data and analysis routines needed to reproduce the figures, as well as the DQMC source code, can be found at \url{https://doi.org/10.5281/zenodo.23582519}. The raw simulation data are available from the corresponding authors upon request.
\clearpage

\begin{figure}[tbp]
    \centering
    \includegraphics[width=\linewidth]{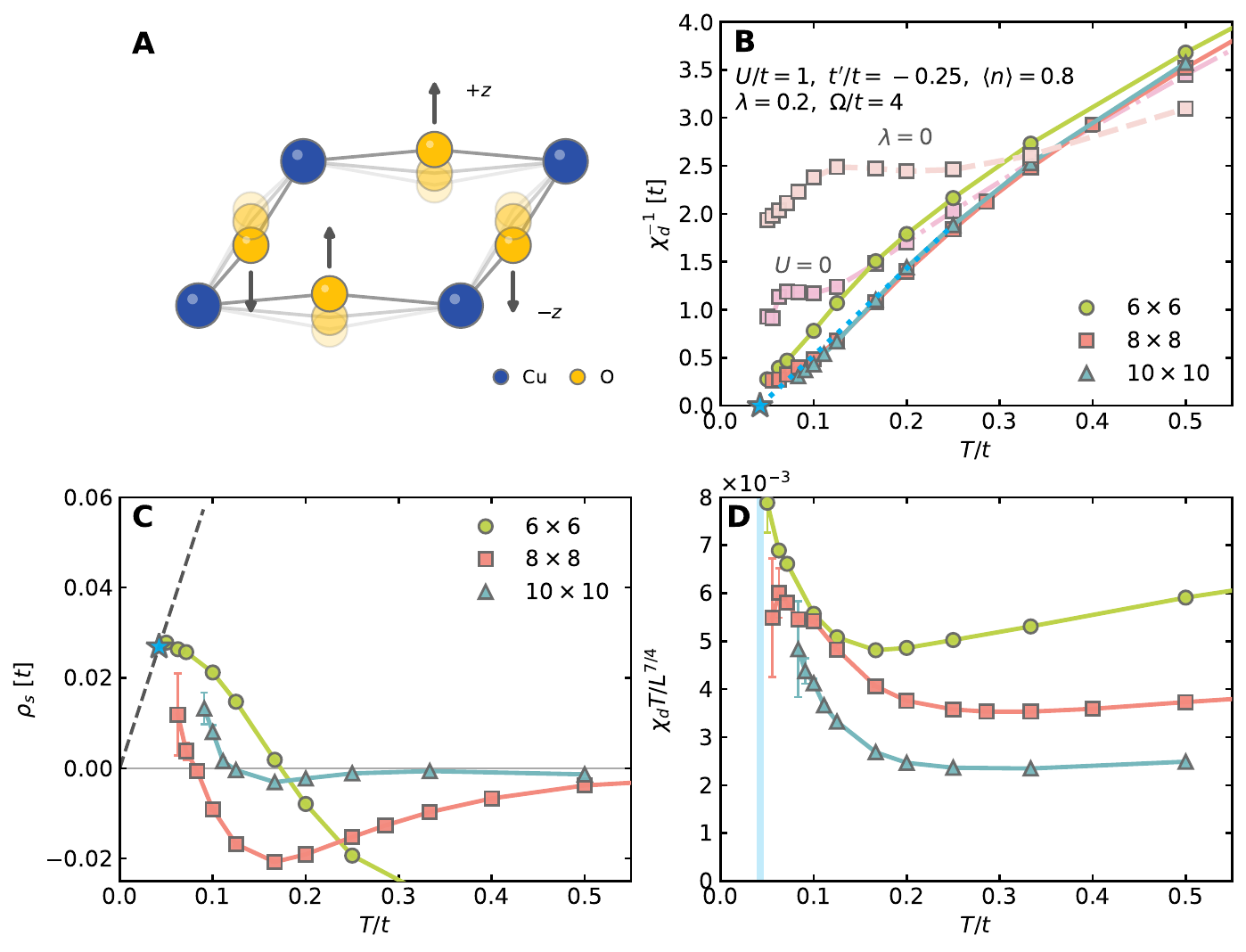}
    \caption{(A) Schematic of the $B_{1g}$ oxygen buckling mode. The planar oxygens on the $x$ and $y$ bonds move out of plane in opposite directions, represented in the model by opposite-sign modulations of the corresponding effective hoppings. (B) Inverse $d$-wave midpoint correlator $\chi_d^{-1}$, where $\chi_d=\beta\langle\mathcal T_\tau\Delta_d(\beta/2)\Delta_d^\dagger(0)\rangle/N_s$. The dash-dotted and dashed curves show $8\times8$ reference results at $U=0$ and $\lambda=0$, respectively. The blue dotted line extrapolates the $10\times10$ data at $U/t=1$; the star marks its zero, the Curie--Weiss temperature $T_0$, which estimates the scale of the superconducting transition temperature. (C) Superfluid density $\rho_s$; the dashed line is the BKT reference $2T/\pi$, and the star marks $T_0$ from (B). (D) $\chi_d T/L^{7/4}$, where $L$ is the linear lattice size; the shaded strip marks the same $T_0$ from (B). Symbols denote lattice size. Parameters: $U/t=1$, $t'/t=-0.25$, $\langle n\rangle=0.8$, $\lambda=0.2$, and $\Omega/t=4$, except for the two reference curves in (B).}
    \label{fig:main_pairing}
\end{figure}

\begin{figure}[tbp]
    \centering
    \includegraphics[width=\linewidth]{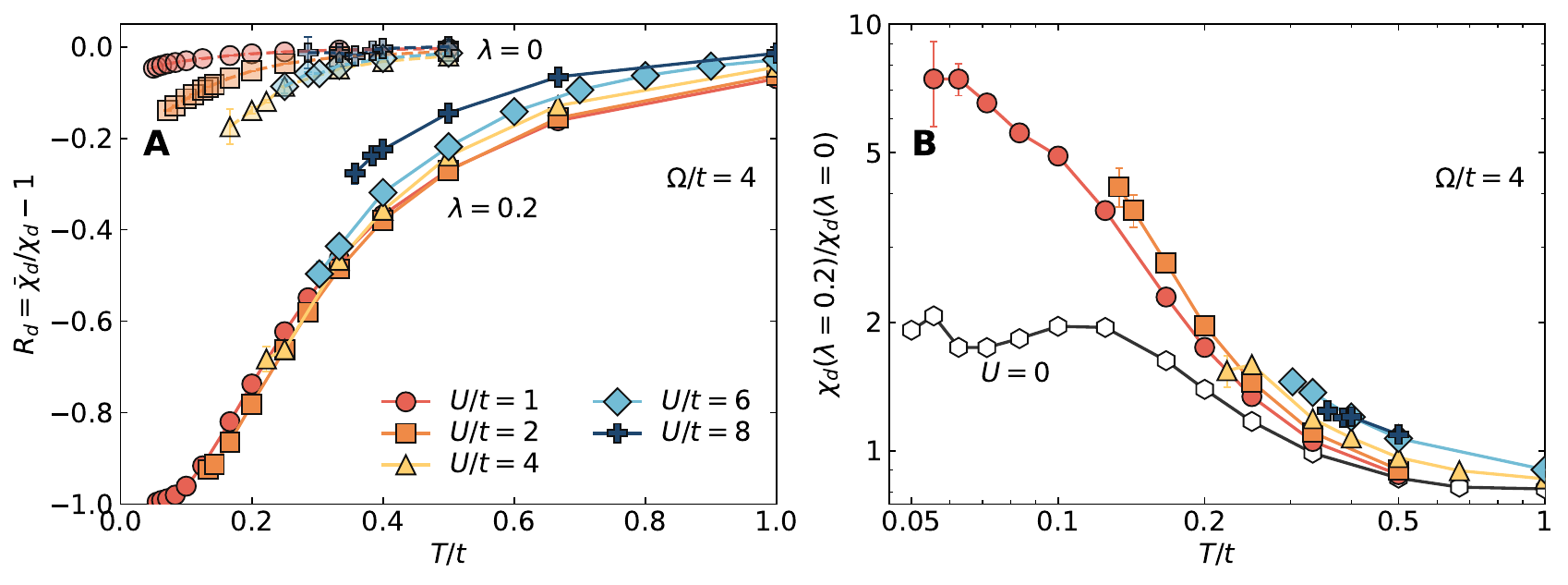}
    \caption{Evolution of $d$-wave pairing with Hubbard repulsion $U$. (A) Vertex ratio $R_d\equiv\bar\chi_d/\chi_d-1$ versus temperature. Solid curves have $\lambda=0.2$; dashed curves are the corresponding $\lambda=0$ Hubbard results. (B) Phonon enhancement $\chi_d(\lambda=0.2)/\chi_d(\lambda=0)$ at the same $U$ and temperature. Parameters: $\Omega/t=4$, $t'/t=-0.25$, $\langle n\rangle=0.8$; lattice size is $8\times8$.}
    \label{fig:main_vertex}
\end{figure}

\begin{figure}[tbp]
    \centering
    \includegraphics[width=\linewidth]{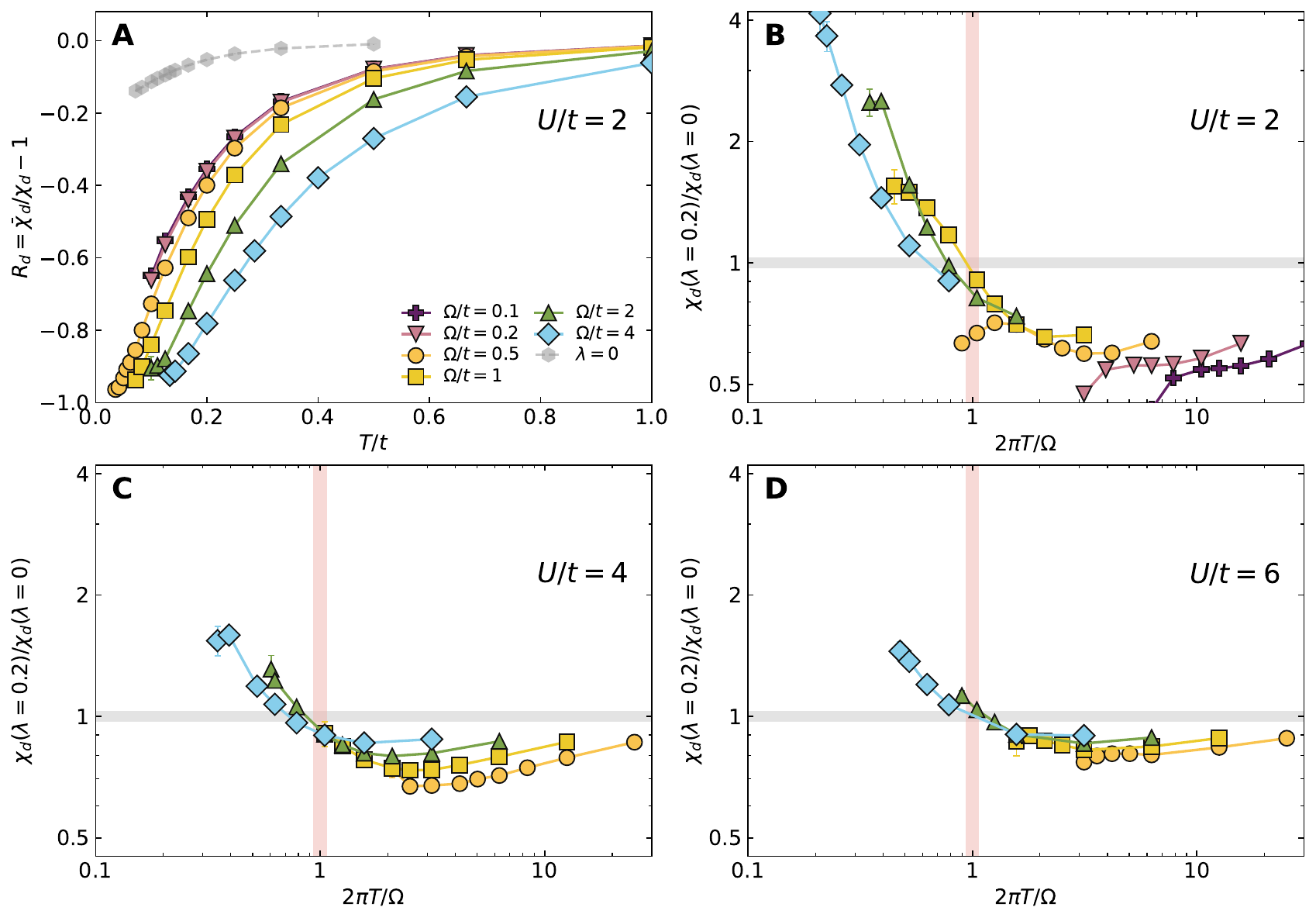}
    \caption{Frequency dependence of $d$-wave pairing. (A) Vertex ratio $R_d=\bar\chi_d/\chi_d-1$ versus temperature at $U/t=2$ for different $\Omega$ and $\lambda=0.2$; the gray dashed curve is the $\lambda=0$ Hubbard result. (B--D) Ratio of midpoint correlators, $\chi_d(\lambda=0.2)/\chi_d(\lambda=0)$, versus $2\pi T/\Omega$ at $U/t=2$ (B), 4 (C), and 6 (D). Colors and symbols distinguish $\Omega$. The gray horizontal band marks unity, and the pink vertical band marks $T=\Omega/(2\pi)$. Parameters: $t'/t=-0.25$, $\langle n\rangle=0.8$; lattice size is $8\times8$.}
    \label{fig:main_omega_crossover}
\end{figure}

\begin{figure}[tbp]
    \centering
    \includegraphics[width=\linewidth]{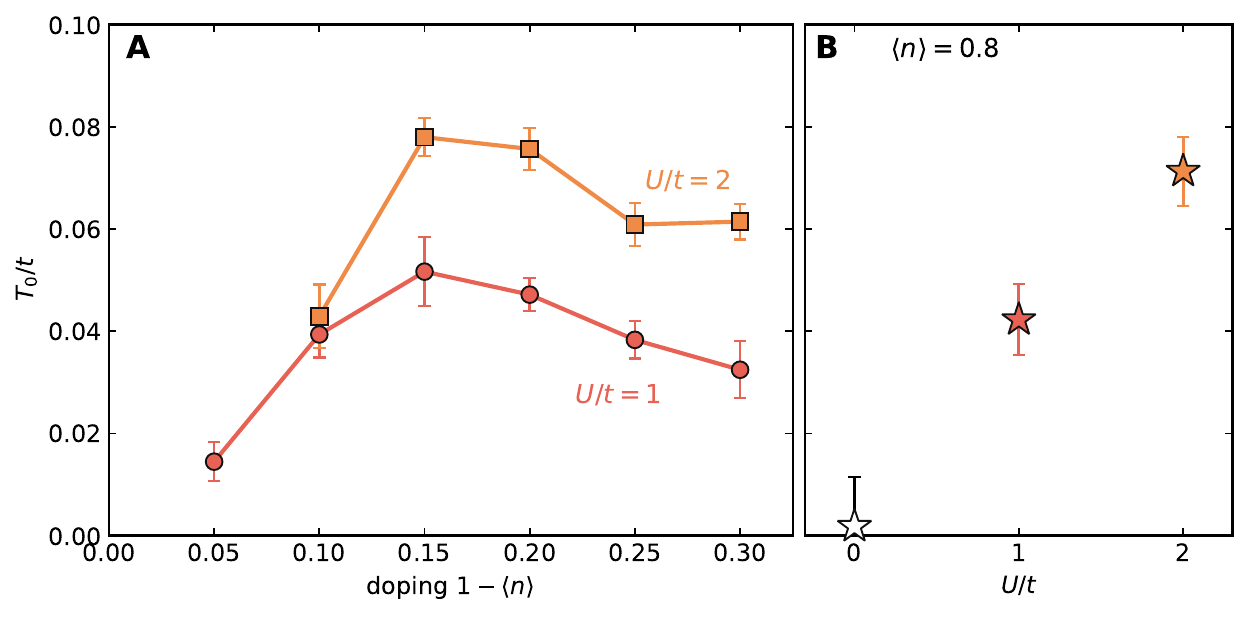}
    \caption{$T_0$ versus doping and Hubbard repulsion.
(A) $T_0/t$ versus doping $1-\langle n\rangle$ at $U/t=1$ and $2$, obtained from Curie--Weiss fits to $8\times8$ data at $\beta t=4,5,6$ (Supplementary Fig.~\ref{fig:supp_u_tc}A,B).
(B) $T_0/t$ versus $U/t$ at $\langle n\rangle=0.8$ from fits to $10\times10$ data at $\beta t=4,5,6$: the leading $A_{1g}$ channel at $U=0$ (open star) and the $d$-wave channel at $U/t=1,2$ (Supplementary Fig.~\ref{fig:supp_u_tc}C,D).
The $U/t=1$ estimates in (A) and (B) agree closely, consistent with size convergence in the Curie--Weiss fitting window.
Parameters: $\lambda=0.2$, $\Omega/t=4$, and $t'/t=-0.25$.}
    \label{fig:main_doping}
\end{figure}

\clearpage
\renewcommand{\thefigure}{S\arabic{figure}}
\renewcommand{\theHfigure}{S\arabic{figure}}
\setcounter{figure}{0}

\section*{Supplementary Materials}
\noindent
Methods\\
Supplementary Text\\
Figs. S1 to S\total{figure}\\
References (\citen{DQMC1}--\total{enumiv})

\clearpage
\section*{Methods}

\subsection*{Simulation details}
We perform determinant quantum Monte Carlo simulations in the grand-canonical ensemble~\cite{DQMC1} on $6\times6$, $8\times8$, and $10\times10$ square lattices with periodic boundary conditions. We set $k_B=\hbar=1$ and discretize imaginary time with $t\Delta\tau\le0.1$.
The Hubbard interaction is decoupled using a discrete Hubbard--Stratonovich transformation~\cite{Hirsch1983HS}. We sample the phonon displacements $Q_i(\tau_\ell)$ as continuous fields, where $\tau_\ell=\ell\Delta\tau$ denotes the $\ell$th imaginary-time slice. For $H_{{\rm ph},i}=P_i^2/(2M)+M\Omega^2Q_i^2/2$, the exact oscillator propagator~\cite{FeynmanHibbs1965} is
\begin{equation*}
\begin{gathered}
\langle Q_i(\tau_{\ell+1})|e^{-\Delta\tau H_{{\rm ph},i}}|Q_i(\tau_\ell)\rangle
=\sqrt{\frac{M\Omega}{2\pi\sinh(\Omega\Delta\tau)}}\\
{}\times\exp\!\left[-\frac{M\Omega\bigl[(Q_i^2(\tau_{\ell+1})+Q_i^2(\tau_\ell))\cosh(\Omega\Delta\tau)-2Q_i(\tau_{\ell+1})Q_i(\tau_\ell)\bigr]}{2\sinh(\Omega\Delta\tau)}\right].
\end{gathered}
\end{equation*}
The harmonic field weight is the product of these propagators over sites and adjacent time slices, with $Q_i(\beta)=Q_i(0)$.

At each time slice, the nearest-neighbor hopping is $t_{ij}(\tau_\ell)=t\pm g[Q_i(\tau_\ell)+Q_j(\tau_\ell)]/2$, with $+$ for $x$ bonds and $-$ for $y$ bonds. With the bond-hopping operator $K_{ij}=\sum_\sigma(c_{i\sigma}^\dagger c_{j\sigma}+c_{j\sigma}^\dagger c_{i\sigma})$, we approximate its kinetic propagator by an ordered product of bond factors,
\begin{equation*}
e^{\Delta\tau\sum_{\langle ij\rangle}t_{ij}(\tau_\ell)K_{ij}}
\simeq\prod_{\langle ij\rangle}^{\rm ordered}
e^{\Delta\tau t_{ij}(\tau_\ell)K_{ij}},
\end{equation*}
grouping nonoverlapping bonds together. For each spin $\sigma$, the bond propagator in the ordered single-particle basis $\bigl(c_{i\sigma}^\dagger|0\rangle,c_{j\sigma}^\dagger|0\rangle\bigr)$ is
\begin{equation*}
\exp\!\left[\Delta\tau t_{ij}(\tau_\ell)
\begin{pmatrix}0&1\\1&0\end{pmatrix}\right]
=
\begin{pmatrix}
\cosh[\Delta\tau t_{ij}(\tau_\ell)] & \sinh[\Delta\tau t_{ij}(\tau_\ell)]\\
\sinh[\Delta\tau t_{ij}(\tau_\ell)] & \cosh[\Delta\tau t_{ij}(\tau_\ell)]
\end{pmatrix}.
\end{equation*}
The $t'$ hopping and chemical potential enter a separate phonon-independent kinetic propagator.

\begin{samepage}
Local Metropolis moves are supplemented by shifts of the whole imaginary-time path at a site~\cite{Johnston2013HubbardHolstein}. We also make collective shifts along real spatial Fourier modes, applying the same displacement pattern at every time slice.
Specifically, we randomly select a normalized sine or cosine spatial mode $f_{\mathbf q}(\mathbf r_i)$ at lattice momentum $\mathbf q$ and propose
\[
Q_i(\tau_\ell)\longrightarrow Q_i(\tau_\ell)+a f_{\mathbf q}(\mathbf r_i),
\]
with a random amplitude $a$. The entire proposal is accepted or rejected in a single Metropolis step.

We also cross-checked our simulations with version 2.0 of the \texttt{SmoQyDQMC.jl} package~\cite{SmoQyDQMC1, SmoQyDQMC2}, which uses an alternative sampling scheme for the phonons based on hybrid Monte Carlo combined with radial and swap updates. 
\end{samepage}

\subsection*{Chemical-potential tuning}
We tune the chemical potential $\mu$ to the target average filling for each parameter set, temperature, and lattice size by interpolating the measured density $\langle n\rangle(\mu)$. The reported fillings denote these target densities.

\subsection*{Error analysis}
Observables are reweighted by the fermion sign.
{
To account for sensitivity to rare, large excursions in individual Markov chains, we combine the jackknife standard error with a spike term:
\[
\sigma=\sqrt{\sigma_{\rm JK}^2+\Delta_{\rm spike}^2}.
\]
Error bars denote $\pm\sigma$.
Here $\sigma_{\rm JK}$ is the jackknife standard error over independent Markov chains $b$ with $n_b$ measurements each,
\[
\sigma_{\rm JK}^2=\frac{1}{\sum_b m_b}\sum_b\frac{m_b^2}{n_b}\big(\theta_{(b)}-\bar\theta\big)^2,
\]
where $\theta$ is the full-sample estimate, $\theta_{(b)}$ is the estimate without chain $b$, $m_b=N-n_b$ with $N=\sum_b n_b$, and $\bar\theta=\sum_b m_b\theta_{(b)}/\sum_b m_b$.
The spike term measures the first-order sensitivity to omitting outlier chains,
\[
\Delta_{\rm spike}=\left|\sum_{b\in\mathcal S}\big(\theta_{(b)}-\theta\big)\right|,
\qquad
\mathcal S=\{b:s_u>0,\ |u_b|>6s_u\},
\]
where $u_b=m_b(\theta_{(b)}-\bar\theta)/\sqrt{n_b}$ and $s_u=1.4826\,\operatorname{med}_{b'}|u_{b'}-\operatorname{med}(u)|$.
} Pairing vertex ratios from the same simulations are evaluated within the same resamples to retain their covariance. Phonon-to-Hubbard pairing ratios are calculated directly from the measured correlators, with uncertainties propagated from the independent simulations. Uncertainties in the non-additivity $S_d$ are propagated from the corresponding uncertainty estimates.

We obtain $T_0$ from Curie--Weiss fits $\chi^{-1}(T)=a(T-T_0)$ with weights $1/\sigma_i^2$, where $\sigma_i$ is the uncertainty in $\chi^{-1}(T_i)$. We fit the $\beta t=4,5,6$ data on $10\times10$ lattices in Figs.~\ref{fig:main_pairing}B and~\ref{fig:main_doping}B and Supplementary Fig.~\ref{fig:supp_estimator}, and on $8\times8$ lattices in Fig.~\ref{fig:main_doping}A, where data are available at every doping studied. At $U=0$ (Fig.~\ref{fig:main_doping}B), the fit uses $\chi_{A_{1g}}^{-1}$ of Eq.~\eqref{eq:a1g_leading} in place of $\chi_d^{-1}$.
We propagate the jackknife replicas, rescaled to reproduce $\sigma$, through the fits to estimate the $T_0$ error and multiply it by $\sqrt{\max(1,\chi_{\rm red}^2)}$, where the reduced chi-square is
\[
\chi_{\rm red}^2
=\frac{1}{3-2}\sum_{i=1}^{3}
\left[\frac{\chi_d^{-1}(T_i)-a(T_i-{T_0})}{\sigma_i}\right]^2.
\]
Here $3-2=1$ is the number of degrees of freedom: three data points minus the two fit parameters $a$ and $T_0$.
This rescaling allows deviations from linearity within the fitting window to enlarge the uncertainty beyond the propagated Monte Carlo error.

\clearpage
\section*{Supplementary Text}

\subsection*{Model and coupling symmetry}

We study the two-dimensional single-band $t$--$t'$--$U$ Hubbard model coupled to local $B_{1g}$ phonons on a square lattice of linear size $L$ and $N_s=L^2$ sites, with periodic boundary conditions in both spatial directions. The Hamiltonian is
\begin{equation}
H=H_{\rm Hub}+H_{\rm ph}+H_{e\text{-}ph}.
\label{eq:b1g_full_model}
\end{equation}
The electronic part is the Hubbard Hamiltonian,
\begin{equation}
H_{\rm Hub}=-t\sum_{\langle ij\rangle}K_{ij}
-t'\sum_{\langle\!\langle ij\rangle\!\rangle}K_{ij}
+U\sum_i n_{i\uparrow}n_{i\downarrow}-\mu\sum_i n_i,
\label{eq:b1g_hubbard_part}
\end{equation}
where $c_{i\sigma}^\dagger$ ($c_{i\sigma}$) creates (annihilates) an electron at site $i$ with spin $\sigma\in\{\uparrow,\downarrow\}$, $n_{i\sigma}=c_{i\sigma}^\dagger c_{i\sigma}$ and $n_i=n_{i\uparrow}+n_{i\downarrow}$ are the spin-resolved and total on-site number operators, and $K_{ij}=\sum_\sigma(c_{i\sigma}^\dagger c_{j\sigma}+c_{j\sigma}^\dagger c_{i\sigma})$ is the bond-hopping operator.
The sums $\langle ij\rangle$ and $\langle\!\langle ij\rangle\!\rangle$ run over nearest-neighbor and next-nearest-neighbor bonds, respectively, with each bond counted once.
The parameters $t$ and $t'$ are the corresponding hopping energies, $U$ is the on-site repulsion, and $\mu$ is the chemical potential controlling the average electron filling $\langle n\rangle=N_s^{-1}\sum_i\langle n_i\rangle$.

The harmonic phonon Hamiltonian is
\begin{equation}
H_{\rm ph}=\sum_i\left(\frac{P_i^2}{2M}+\frac{M\Omega^2Q_i^2}{2}\right).
\label{eq:b1g_phonon_part}
\end{equation}
The coordinate $Q_i$ describes the local $B_{1g}$ buckling motion of the planar oxygens surrounding site $i$; $P_i$ is its conjugate momentum, $M$ the effective phonon mass, and $\Omega$ the bare phonon frequency.
The electron--phonon coupling is
\begin{equation}
\begin{aligned}
H_{e\text{-}ph}&=-g\sum_i Q_i\mathcal K_{B_{1g},i},\\
\mathcal K_{B_{1g},i}&=\tfrac12\bigl(K_{i,i+\hat x}+K_{i,i-\hat x}
-K_{i,i+\hat y}-K_{i,i-\hat y}\bigr),
\end{aligned}
\label{eq:b1g_electron_phonon}
\end{equation}
where $g$ is the linear coupling strength and $\hat x$ and $\hat y$ are unit vectors along the two lattice directions.
We set $\hbar=k_B=1$ and the lattice constant to unity, and use $t$ as the energy unit, with inverse temperature $\beta=1/T$.
We take $t'=-0.25t$ and primarily study $\langle n\rangle=0.8$.
The dimensionless coupling is $\lambda=g^2/(2Mt\Omega^2)$; we use $\lambda=0.2$ for calculations with phonons and $\lambda=0$ for the pure Hubbard reference results.

The $B_{1g}$ symmetry requires opposite hopping modulations along the two planar directions. For the positive-direction bonds, Eq.~\eqref{eq:b1g_electron_phonon} gives
\begin{equation}
\delta t_{i,i+\hat x}=\frac{g}{2}(Q_i+Q_{i+\hat x}),
\qquad
\delta t_{i,i+\hat y}=-\frac{g}{2}(Q_i+Q_{i+\hat y}),
\label{eq:b1g_bond_modulations}
\end{equation}
so that $H_{e\text{-}ph}=-\sum_{i,\alpha=x,y}\delta t_{i,i+\hat\alpha}K_{i,i+\hat\alpha}$. The single coordinate $Q_i$ correlates the four bonds around site $i$.

We denote the position of site $i$ by $\mathbf r_i$ and the electron and phonon momenta by $\mathbf k$ and $\mathbf q$, respectively.
Using $c_{i\sigma}=N_s^{-1/2}\sum_{\mathbf k}e^{i\mathbf k\cdot\mathbf r_i}c_{\mathbf k\sigma}$ and $Q_i=N_s^{-1/2}\sum_{\mathbf q}e^{i\mathbf q\cdot\mathbf r_i}Q_{\mathbf q}$, the coupling becomes
\begin{equation}
\begin{aligned}
H_{e\text{-}ph}&=-\frac{1}{\sqrt{N_s}}\sum_{\mathbf k,\mathbf q,\sigma}
Q_{\mathbf q}\Lambda_{B_{1g}}(\mathbf k,\mathbf q)
c_{\mathbf k+\mathbf q,\sigma}^\dagger c_{\mathbf k\sigma},\\
\Lambda_{B_{1g}}(\mathbf k,\mathbf q)
&=2g\left[\cos\left(k_x+\frac{q_x}{2}\right)\cos\frac{q_x}{2}
-\cos\left(k_y+\frac{q_y}{2}\right)\cos\frac{q_y}{2}\right]\\
&=g\bigl[f_d(\mathbf k)+f_d(\mathbf k+\mathbf q)\bigr].
\end{aligned}
\label{eq:b1g_momentum_vertex}
\end{equation}
Here $f_d(\mathbf k)=\cos k_x-\cos k_y$ is the $d_{x^2-y^2}$-wave form factor. For Cooper scattering $(\mathbf k,-\mathbf k)\to(\mathbf k',-\mathbf k')$, the momentum transfer is $\mathbf q=\mathbf k'-\mathbf k$, giving $\Lambda_{B_{1g}}(\mathbf k,\mathbf k'-\mathbf k)=g[f_d(\mathbf k)+f_d(\mathbf k')]$. For a uniform buckling coordinate, $\mathbf q=0$ and $\Lambda_{B_{1g}}(\mathbf k,\mathbf 0)=2g f_d(\mathbf k)$.

At fixed electron fields, the phonon part of the Euclidean path integral in imaginary time $\tau\in[0,\beta]$ starts from
\begin{equation}
\int\!\left[\prod_i\mathcal DQ_i\mathcal DP_i\right]
\exp\!\left\{-\sum_i\int_0^\beta d\tau\left[
iP_i\partial_\tau Q_i+\frac{P_i^2}{2M}
+\frac{M\Omega^2Q_i^2}{2}-gQ_i\mathcal K_{B_{1g},i}
\right]\right\}.
\label{eq:b1g_phase_space_integral}
\end{equation}
Completing $P_i^2/(2M)+iP_i\partial_\tau Q_i=(P_i+iM\partial_\tau Q_i)^2/(2M)+M(\partial_\tau Q_i)^2/2$, the $P_i$ integral contributes an electron-independent constant and leaves
\begin{equation}
\int\!\left[\prod_i\mathcal DQ_i\right]e^{-S_Q},\qquad
S_Q=\sum_i\int_0^\beta d\tau\left[
\frac{M}{2}(\partial_\tau Q_i)^2
+\frac{M\Omega^2}{2}Q_i^2
-gQ_i\mathcal K_{B_{1g},i}\right].
\label{eq:b1g_phonon_integral}
\end{equation}
With $Q_i(i\nu_n)=\beta^{-1/2}\int_0^\beta d\tau\,e^{i\nu_n\tau}Q_i(\tau)$ and the same convention for $\mathcal K_{B_{1g},i}$, where $\nu_n=2\pi n/\beta$ with $n\in\mathbb Z$, this action becomes
\begin{equation}
S_Q=\sum_{i,\nu_n}\left[
\frac{M(\nu_n^2+\Omega^2)}{2}Q_i(-i\nu_n)Q_i(i\nu_n)
-gQ_i(-i\nu_n)\mathcal K_{B_{1g},i}(i\nu_n)\right].
\label{eq:b1g_phonon_frequency_action}
\end{equation}
Defining $\widetilde Q_i(i\nu_n)=Q_i(i\nu_n)-g\mathcal K_{B_{1g},i}(i\nu_n)/[M(\nu_n^2+\Omega^2)]$ completes the square:
\begin{equation}
\begin{aligned}
S_Q&=\frac{M}{2}\sum_{i,\nu_n}(\nu_n^2+\Omega^2)
\widetilde Q_i(-i\nu_n)\widetilde Q_i(i\nu_n)+S_{\rm ind},\\
S_{\rm ind}&=-\frac{g^2}{2M}\sum_{i,\nu_n}
\frac{\mathcal K_{B_{1g},i}(-i\nu_n)\mathcal K_{B_{1g},i}(i\nu_n)}
{\nu_n^2+\Omega^2}.
\end{aligned}
\label{eq:b1g_induced_action}
\end{equation}
The Gaussian integral over $\widetilde Q_i$ is electron-independent, hence $\int[\prod_i\mathcal DQ_i]e^{-S_Q}=C e^{-S_{\rm ind}}$, with constant $C$. To extract the opposite-spin pairing interaction, separate $\mathcal K_{B_{1g},i}=\mathcal K_{B_{1g},i\uparrow}+\mathcal K_{B_{1g},i\downarrow}$ and Fourier transform its spin-resolved terms:
\begin{equation}
\mathcal K_{B_{1g},\sigma}(\mathbf q,\tau)
=\sum_i e^{i\mathbf q\cdot\mathbf r_i}\mathcal K_{B_{1g},i\sigma}(\tau)
=\sum_{\mathbf k}\bigl[f_d(\mathbf k)+f_d(\mathbf k+\mathbf q)\bigr]
c_{\mathbf k+\mathbf q,\sigma}^\dagger(\tau)c_{\mathbf k\sigma}(\tau).
\label{eq:b1g_spin_bilinear}
\end{equation}
The two cross terms in $S_{\rm ind}$ are equal under $(\mathbf q,\nu_n)\to(-\mathbf q,-\nu_n)$, canceling its factor $1/2$. Using $\sum_i e^{i(\mathbf q-\mathbf q')\cdot\mathbf r_i}=N_s\delta_{\mathbf q,\mathbf q'}$ and the same time-Fourier convention as above, they give
\begin{equation}
S_{\rm ind}^{\uparrow\downarrow}
=-\frac{g^2}{MN_s}\sum_{\mathbf q,\nu_n}
\frac{\mathcal K_{B_{1g},\uparrow}(\mathbf q,-i\nu_n)
\mathcal K_{B_{1g},\downarrow}(-\mathbf q,i\nu_n)}
{\nu_n^2+\Omega^2}.
\label{eq:b1g_opposite_spin_action}
\end{equation}
For Cooper scattering $(\mathbf k,-\mathbf k)\to(\mathbf k',-\mathbf k')$, set $\mathbf q=\mathbf k'-\mathbf k$ in Eq.~\eqref{eq:b1g_opposite_spin_action}. The two form factors in Eq.~\eqref{eq:b1g_spin_bilinear}, together with the factor $g^2$, give the product of vertices below. Reordering the fermion fields into pair creation and annihilation operators gives the retarded phonon-mediated pairing interaction
\begin{equation}
\begin{aligned}
V_{\rm ph}(\mathbf k,\mathbf k';i\nu_n)
&=-\frac{\Lambda_{B_{1g}}(\mathbf k,\mathbf k'-\mathbf k)
\Lambda_{B_{1g}}(-\mathbf k,\mathbf k-\mathbf k')}
{M(\nu_n^2+\Omega^2)}\\
&=-\frac{g^2[f_d(\mathbf k)+f_d(\mathbf k')]^2}
{M(\nu_n^2+\Omega^2)}.
\end{aligned}
\label{eq:b1g_retarded_interaction}
\end{equation}
Adding the Hubbard repulsion gives the bare pairing interaction,
\begin{equation}
\begin{aligned}
V_{\rm pair}(\mathbf k,\mathbf k';i\nu_n)
&=U+V_{\rm ph}(\mathbf k,\mathbf k';i\nu_n)\\
&=U-\frac{2\lambda t\Omega^2}{\nu_n^2+\Omega^2}
\bigl[f_d(\mathbf k)^2+f_d(\mathbf k')^2
+2f_d(\mathbf k)f_d(\mathbf k')\bigr].
\end{aligned}
\label{eq:b1g_pair_channels}
\end{equation}
The on-site $s$-wave form factor $1$ and $f_d^2$ are distinct $A_{1g}$ functions, although $f_d^2$ contains a constant component. In Eq.~\eqref{eq:b1g_pair_channels}, $f_d(\mathbf k)^2$ and $f_d(\mathbf k')^2$ implicitly multiply $1(\mathbf k')$ and $1(\mathbf k)$, respectively: the phonon mixes these two $A_{1g}$ form factors, while $U$ contributes the explicit $1(\mathbf k)1(\mathbf k')$ repulsion. The $2f_d(\mathbf k)f_d(\mathbf k')$ term provides direct $d$-wave attraction.

\subsection*{Formalism}

We define a singlet on the bond from site $i$ to $i+\hat\alpha$ by
\begin{equation}
\Delta_{\hat\alpha}^\dagger(i)=\frac{1}{\sqrt{2}}
\bigl(c_{i\uparrow}^\dagger c_{i+\hat\alpha,\downarrow}^\dagger
-c_{i\downarrow}^\dagger c_{i+\hat\alpha,\uparrow}^\dagger\bigr),
\qquad \hat\alpha\in\{\hat x,\hat y\}.
\end{equation}
The zero-momentum $d$-wave operator is
$\Delta_d^\dagger=2^{-1/2}\sum_i
(\Delta_{\hat x}^\dagger(i)-\Delta_{\hat y}^\dagger(i))$,
with momentum form factor $f_d(\mathbf k)=\cos k_x-\cos k_y$.
For comparison, we use two $A_{1g}$ pairing operators: the on-site $s$-wave operator
$\Delta_s^\dagger=\sum_i c_{i\uparrow}^\dagger c_{i\downarrow}^\dagger$
and the $f_d^2$-weighted operator $\Delta_{f_d^2}^\dagger=\sum_{\mathbf k}f_d(\mathbf k)^2
c_{\mathbf k\uparrow}^\dagger c_{-\mathbf k\downarrow}^\dagger$.

For each pairing operator $\alpha\in\{s,d,f_d^2\}$, the midpoint correlator is
\begin{equation}
\chi_\alpha=\frac{\beta}{N_s}
\langle\mathcal T_\tau\Delta_\alpha(\beta/2)\Delta_\alpha^\dagger(0)\rangle.
\label{eq:pair_midpoint}
\end{equation}
Here $N_s$ is the number of lattice sites, and $\chi_\alpha$ has units of $t^{-1}$.
For the bond correlator
$S_{\hat\alpha,\hat\alpha'}(\mathbf r,\tau)=N_s^{-1}\sum_i\langle\mathcal T_\tau\Delta_{\hat\alpha}(\tau,\mathbf r_i+\mathbf r)\Delta_{\hat\alpha'}^\dagger(0,\mathbf r_i)\rangle$
(also defined in Supplementary Fig.~\ref{fig:supp_realspace}), this normalization gives
\begin{equation}
\chi_d=\frac{\beta}{2}\sum_{\mathbf r}
\bigl[S_{\hat x,\hat x}+S_{\hat y,\hat y}
-S_{\hat x,\hat y}-S_{\hat y,\hat x}\bigr](\mathbf r,\beta/2).
\end{equation}
The bare bubble is obtained by Wick contraction using the sign-reweighted
single-particle Green's functions $G_\sigma$:
\begin{equation}
\bar\chi_d=\frac{\beta}{N_s}\sum_{\mathbf k}f_d(\mathbf k)^2
G_\uparrow(\mathbf k,\beta/2)G_\downarrow(-\mathbf k,\beta/2).
\label{eq:pair_bare_bubble}
\end{equation}
We compare it with the full correlator through
$R_d=\bar\chi_d/\chi_d-1$; $R_d<0$ indicates a $d$-wave vertex enhancement.
Further details of the midpoint and bond-correlator formalism are given in
Ref.~\cite{Wang2025FiniteTemperature}.

For $\alpha=s$ and $f_d^2$, the corresponding midpoint correlators are
\begin{equation}
\begin{aligned}
\chi_s&=\frac{\beta}{N_s}\sum_{i,j}
\langle\mathcal T_\tau c_{i\downarrow}(\beta/2)c_{i\uparrow}(\beta/2)
c_{j\uparrow}^\dagger(0)c_{j\downarrow}^\dagger(0)\rangle,\\
\chi_{f_d^2}&=\frac{\beta}{N_s}
\langle\mathcal T_\tau\Delta_{f_d^2}(\beta/2)\Delta_{f_d^2}^\dagger(0)\rangle.
\end{aligned}
\label{eq:a1g_midpoint}
\end{equation}
Their bare bubbles follow Eq.~\eqref{eq:pair_bare_bubble} with the weight $f_d(\mathbf k)^2$ replaced by $1$ and $f_d(\mathbf k)^4$, respectively. We define $R_\alpha=\bar\chi_\alpha/\chi_\alpha-1$ for these channels in the same way.

We calculate the superfluid density $\rho_s$ by averaging the $x$- and $y$-direction longitudinal-minus-transverse current estimators within each chain~\cite{PhysRevLett.68.2830,Wang2025FiniteTemperature}:
\begin{equation}
\begin{aligned}
\rho_s&=\frac{1}{8}(\Lambda_{xx}^{L}-\Lambda_{xx}^{T}+\Lambda_{yy}^{L}-\Lambda_{yy}^{T}),\\
\Lambda_{xx}^{L}&=\lim_{q_x\to0}\Lambda_{xx}((q_x,0),i\nu_0),\\
\Lambda_{xx}^{T}&=\lim_{q_y\to0}\Lambda_{xx}((0,q_y),i\nu_0).
\end{aligned}
\label{eq:superfluid_density}
\end{equation}
The response uses the phonon-dressed bond current. For a bond $b$ oriented from $i_b$ to $j_b$, the Peierls substitution and its current operator are
\begin{equation}
\begin{aligned}
H_{\rm kin}(\mathbf A)&=-\sum_{b,\sigma}t_b(Q)\bigl[e^{i\mathbf A(\mathbf R_b)\cdot\mathbf d_b}c_{j_b\sigma}^\dagger c_{i_b\sigma}+\mathrm{H.c.}\bigr],\\
j_{x,b}(\tau)&=i d_{b,x}t_b[Q(\tau)]\sum_\sigma
\bigl(c_{j_b\sigma}^\dagger(\tau)c_{i_b\sigma}(\tau)-\mathrm{H.c.}\bigr),
\end{aligned}
\label{eq:phonon_dressed_current}
\end{equation}
where $\mathbf d_b=\mathbf r_{j_b}-\mathbf r_{i_b}$, $\mathbf R_b=(\mathbf r_{i_b}+\mathbf r_{j_b})/2$, and the sum includes each nearest- and next-nearest-neighbor bond once. In the present model,
\begin{equation}
\begin{aligned}
t_{i,i+\hat x}(Q)&=t+\frac{g}{2}(Q_i+Q_{i+\hat x}),\\
t_{i,i+\hat y}(Q)&=t-\frac{g}{2}(Q_i+Q_{i+\hat y}),\\
t_{ij}(Q)&=t'\quad\text{for next-nearest neighbors}.
\end{aligned}
\label{eq:phonon_dressed_hoppings}
\end{equation}
Thus the current includes the instantaneous phonon modulation of the nearest-neighbor hopping at each imaginary-time insertion. With $\nu_n=2\pi nT$, the current response is
\begin{equation}
\Lambda_{xx}(\mathbf q,i\nu_n)=\frac{1}{N_s}\sum_{b,c}e^{-i\mathbf q\cdot(\mathbf R_b-\mathbf R_c)}
\int_0^\beta d\tau\,e^{i\nu_n\tau}
\langle\mathcal T_\tau j_{x,b}(\tau)j_{x,c}(0)\rangle.
\label{eq:superfluid_current_response}
\end{equation}
On a finite $L\times L$ lattice, we evaluate the static longitudinal and transverse responses at $(2\pi/L,0)$ and $(0,2\pi/L)$, respectively. The $y$-direction responses follow by exchanging $x$ and $y$.

\subsection*{Supplementary Figures}

\begin{figure}[!htbp]
    \centering
    \includegraphics[height=0.64\textheight,keepaspectratio]{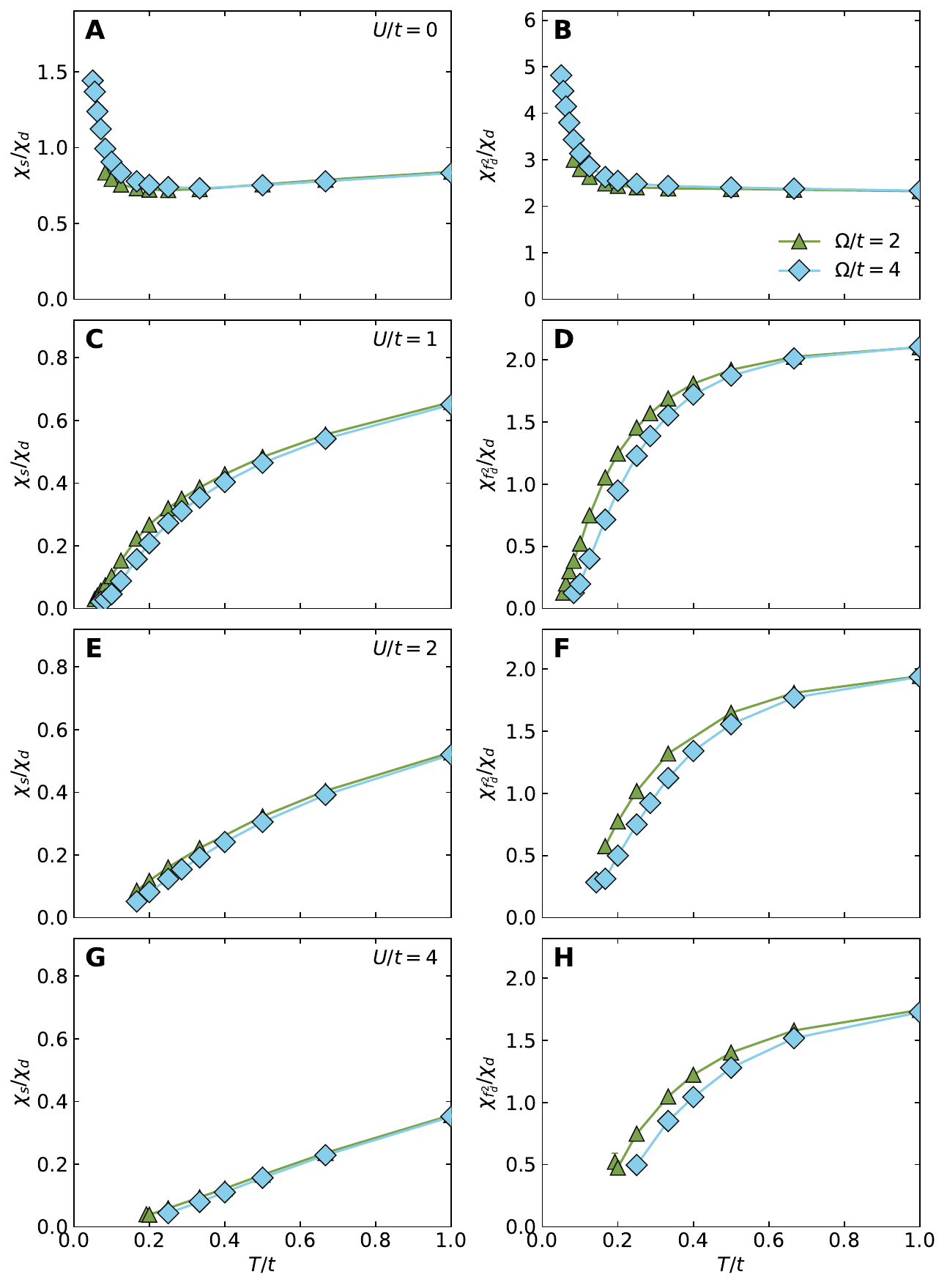}
    \caption{Ratios of the on-site $s$-wave midpoint correlator (left column) and the $f_d^2$-weighted $A_{1g}$ midpoint correlator (right column) to the $d$-wave result, $\chi_s/\chi_d$ and $\chi_{f_d^2}/\chi_d$, with $f_d(\mathbf k)=\cos k_x-\cos k_y$. (A,B) $U/t=0$; (C,D) $U/t=1$; (E,F) $U/t=2$; (G,H) $U/t=4$. Symbols distinguish $\Omega/t=2,4$. Parameters: $t'/t=-0.25$, $\lambda=0.2$, $\langle n\rangle=0.8$; lattice size is $8\times8$.}
    \label{fig:supp_a1g}
\end{figure}

Supplementary Fig.~\ref{fig:supp_a1g} shows the ratios $\chi_s/\chi_d$ and $\chi_{f_d^2}/\chi_d$. Both grow upon cooling at $U=0$, indicating an increasing relative weight of these $A_{1g}$ pairing operators. For $U/t=1,2,4$, both fall rapidly instead: weak Hubbard repulsion already suppresses their correlations relative to $d$-wave pairing.

To estimate the crossover interaction $U_c$ at which the $d$-wave vertex enhancement overtakes the competing $A_{1g}$ channels, we examine
\begin{equation}
\frac{\chi_X/\bar\chi_X}{\chi_d/\bar\chi_d}
=\frac{1+R_d}{1+R_X},
\qquad X\in\{s,f_d^2\},
\label{eq:relative_vertex_enhancement}
\end{equation}
where $X=s$ denotes the on-site $s$-wave channel and $X=f_d^2$ the $f_d^2$-weighted $A_{1g}$ channel. A ratio above (below) unity indicates stronger vertex enhancement in $X$ ($d$), as shown in Supplementary Fig.~\ref{fig:supp_uc_vertex}. The crossing through unity occurs near the reversal of the temperature trend: on the $A_{1g}$ side, the relative enhancement grows upon cooling, whereas on the $d$-wave side it decreases.

For $\Omega/t=2,4$, the on-site $s$-wave channel crosses at slightly smaller $U$ than the $f_d^2$ channel at low temperature, with crossings in both channels at $U_c/t\sim0.1$. Thus even a small Hubbard repulsion suppresses the phonon's relative enhancement of these $A_{1g}$ channels and favors $d$-wave pairing. This small scale explains why $U/t=1$ can suppress the competing $A_{1g}$ enhancement while keeping the fermion sign manageable, allowing us to resolve the $d$-wave superconducting transition in Fig.~\ref{fig:main_pairing} of the main text.

\begin{figure}[tbp]
    \centering
    \includegraphics[width=\linewidth]{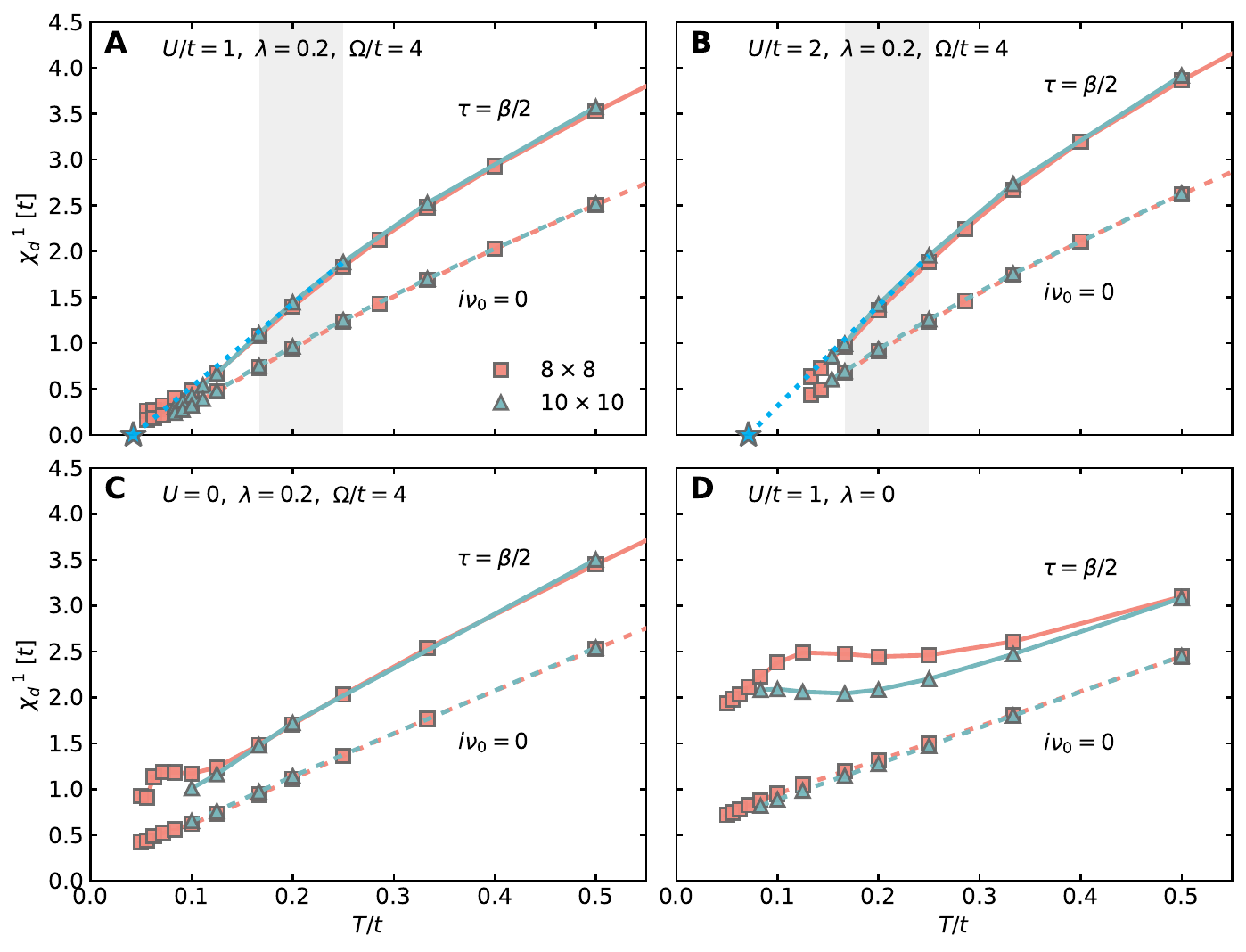}
    \caption{Midpoint and static $d$-wave pairing susceptibilities. $\chi_d^{-1}$ from the midpoint correlator $\chi_d$ of Eq.~\eqref{eq:pair_midpoint} (solid, $\tau=\beta/2$) and from the static susceptibility $\chi_d(i\nu_0=0)$ of Eq.~\eqref{eq:pair_static} (dashed) on $8\times8$ and $10\times10$ lattices: (A) $U/t=1$ and (B) $U/t=2$ at $\lambda=0.2$, $\Omega/t=4$; (C) $U=0$, $\lambda=0.2$, $\Omega/t=4$; (D) the Hubbard model at $U/t=1$, $\lambda=0$. In (A,B), the dotted line is the Curie--Weiss fit to the $10\times10$ midpoint data at $\beta t=4,5,6$ (gray band), and the star marks its zero $T_0$. Parameters: $t'/t=-0.25$, $\langle n\rangle=0.8$.}
    \label{fig:supp_estimator}
\end{figure}

Supplementary Fig.~\ref{fig:supp_estimator} compares the midpoint correlator $\chi_d$ of Eq.~\eqref{eq:pair_midpoint} with the static pairing susceptibility conventionally used in DQMC to probe superconducting order,
\begin{equation}
\chi_d(i\nu_0=0)=\frac{1}{N_s}\int_0^\beta d\tau\,\langle\mathcal T_\tau\Delta_d(\tau)\Delta_d^\dagger(0)\rangle.
\label{eq:pair_static}
\end{equation}
In the superconducting cases at $U/t=1,2$ (A,B), both inverse susceptibilities decrease upon cooling.
The midpoint inverse susceptibility is larger at relatively high temperatures and exhibits a stronger temperature dependence.
In the temperature window where finite-size effects are minimal, Curie--Weiss fits to the two probes yield positive, mutually consistent values of $T_0$.
Upon cooling toward the transition, the two inverse susceptibilities approach quantitative agreement.

In contrast, the controls (C,D) show no comparable trend toward a finite positive $T_0$ over the accessible temperature range, and the two inverse susceptibilities do not approach one another upon cooling.
This behavior is consistent with either the absence of a $d$-wave transition or a transition temperature far below those accessed here.
The midpoint response therefore distinguishes the superconducting cases from these controls more clearly at relatively high temperatures.
Finite-size effects are also generally more pronounced in the midpoint response, reflecting its greater sensitivity to low-energy physics.

\begin{figure}[tbp]
    \centering
    \includegraphics[width=\linewidth]{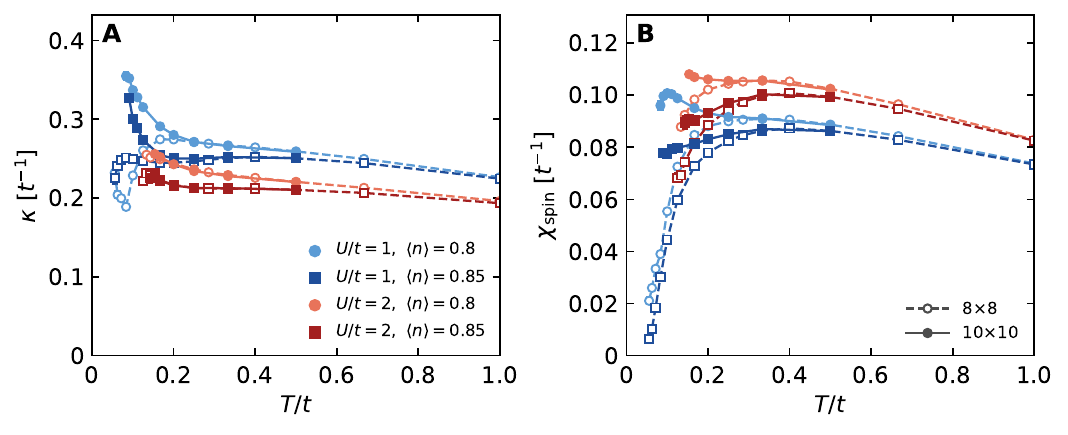}
    \caption{(A) Charge compressibility $\kappa=\beta[\langle N^2\rangle-\langle N\rangle^2]/N_s$ and (B) uniform spin susceptibility $\chi_{\rm spin}=\beta[\langle(S^z_{\rm tot})^2\rangle-\langle S^z_{\rm tot}\rangle^2]/N_s$ versus temperature for $U/t=1,2$ and $\langle n\rangle=0.8,0.85$. Open symbols and dashed curves denote $8\times8$; filled symbols and solid curves denote $10\times10$. Parameters: $t'/t=-0.25$, $\lambda=0.2$, $\Omega/t=4$.}
    \label{fig:supp_thermodynamics}
\end{figure}

We also note that the low-energy thermodynamic responses---the charge compressibility $\kappa=\partial\langle n\rangle/\partial\mu$ and the uniform spin susceptibility $\chi_{\rm spin}=\left.\partial m/\partial h\right|_{h=0}$---show signatures expected of a spin-singlet superconducting state at low temperatures (Supplementary Fig.~\ref{fig:supp_thermodynamics}).
Here $m=\langle S^z_{\rm tot}\rangle/N_s$ is the magnetization density induced by a uniform external Zeeman field $h$ coupled through $-hS^z_{\rm tot}$.

At high temperatures, the two responses track each other and vary only weakly with temperature, as expected for a Fermi-liquid-like response deep in the normal state.
Upon cooling toward $T_0$, their temperature dependences separate: despite finite-size effects at the lowest temperatures, $\kappa$ stays finite over the accessible temperature range, whereas $\chi_{\rm spin}$ shows signs of a downturn at some dopings and lattice sizes.
This contrast is consistent with the development of singlet pairing, which suppresses low-energy spin excitations~\cite{PhysRev.110.769} while leaving collective charge excitations gapless.
These thermodynamic responses provide independent support for the spin-singlet superconducting transition identified in Fig.~\ref{fig:main_pairing} of the main text.
The convergence of pairing, superfluid, and thermodynamic signatures on the same low-temperature superconducting state is particularly notable in a repulsive Hubbard model with a fermion sign problem, where identifying superconductivity in unbiased finite-temperature simulations has historically been extremely difficult.

\begin{figure}[tbp]
    \centering
    \includegraphics[width=\linewidth]{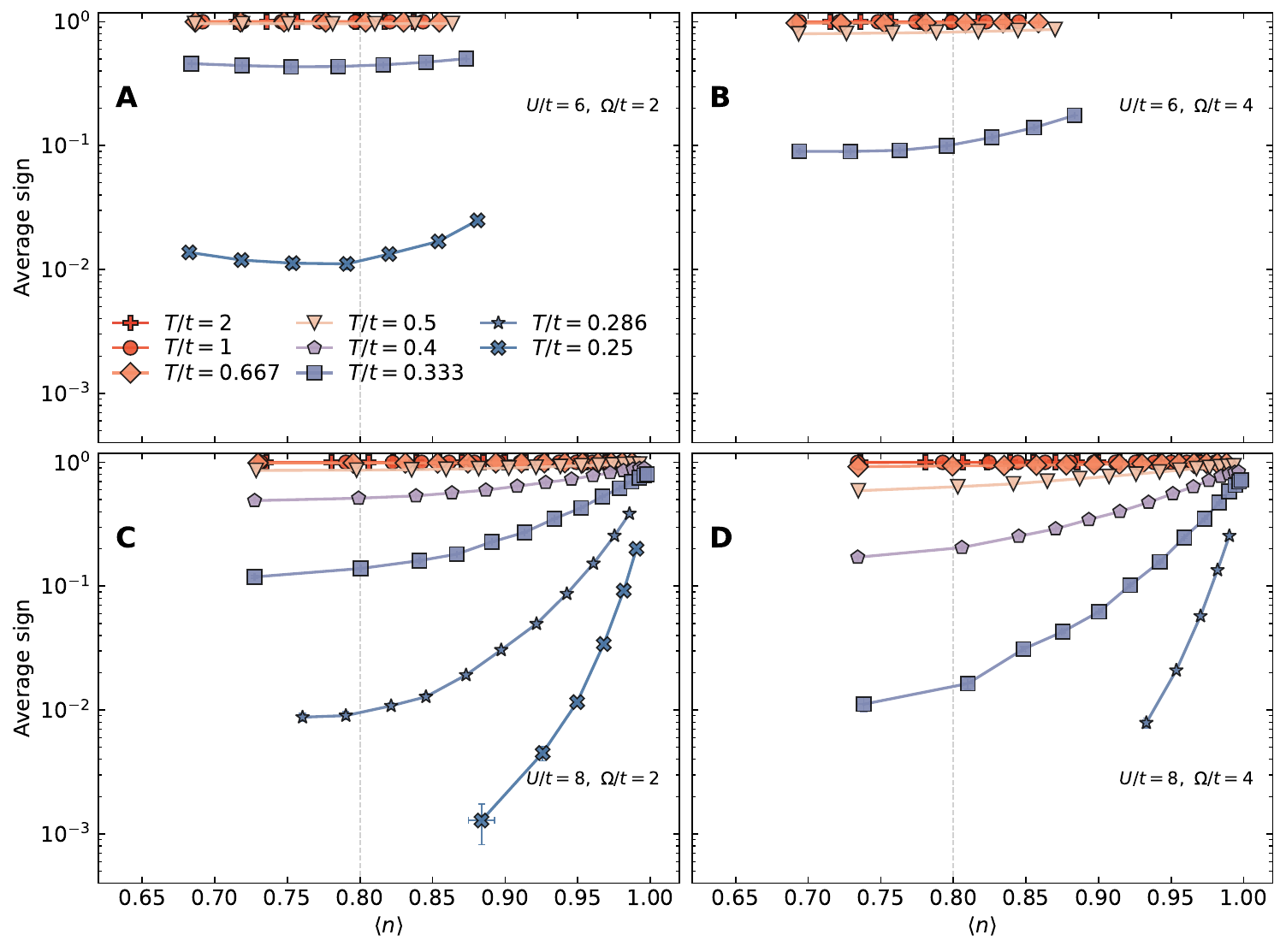}
    \caption{Average fermion sign versus measured filling for $U/t=6$ (A,B) and 8 (C,D), with $\Omega/t=2$ (A,C) and 4 (B,D). Colors and symbols distinguish temperature; the dashed vertical line marks $\langle n\rangle=0.8$. Parameters: $t'/t=-0.25$, $\lambda=0.2$; lattice size is $8\times8$. The sign axis is logarithmic.}
    \label{fig:supp_sign_u68}
\end{figure}

At weak $U$, the manageable fermion sign lets us reach temperatures with superconducting signatures. Supplementary Fig.~\ref{fig:supp_sign_u68} shows why this is harder at $U/t=6,8$: the sign falls rapidly on cooling, leaving their lower-temperature behavior unresolved. At fixed temperature, smaller $\Omega$ generally gives a larger average sign, as shown in Supplementary Fig.~\ref{fig:supp_sign_temperature}A--E. On the $\Omega/(2\pi T)$ scale, however, the sign falls at smaller scaled inverse temperatures for smaller $\Omega$, as shown in Supplementary Fig.~\ref{fig:supp_sign_temperature}F--J, making low $T/\Omega$ harder to access.

\begin{figure}[tbp]
    \centering
    \includegraphics[width=\linewidth]{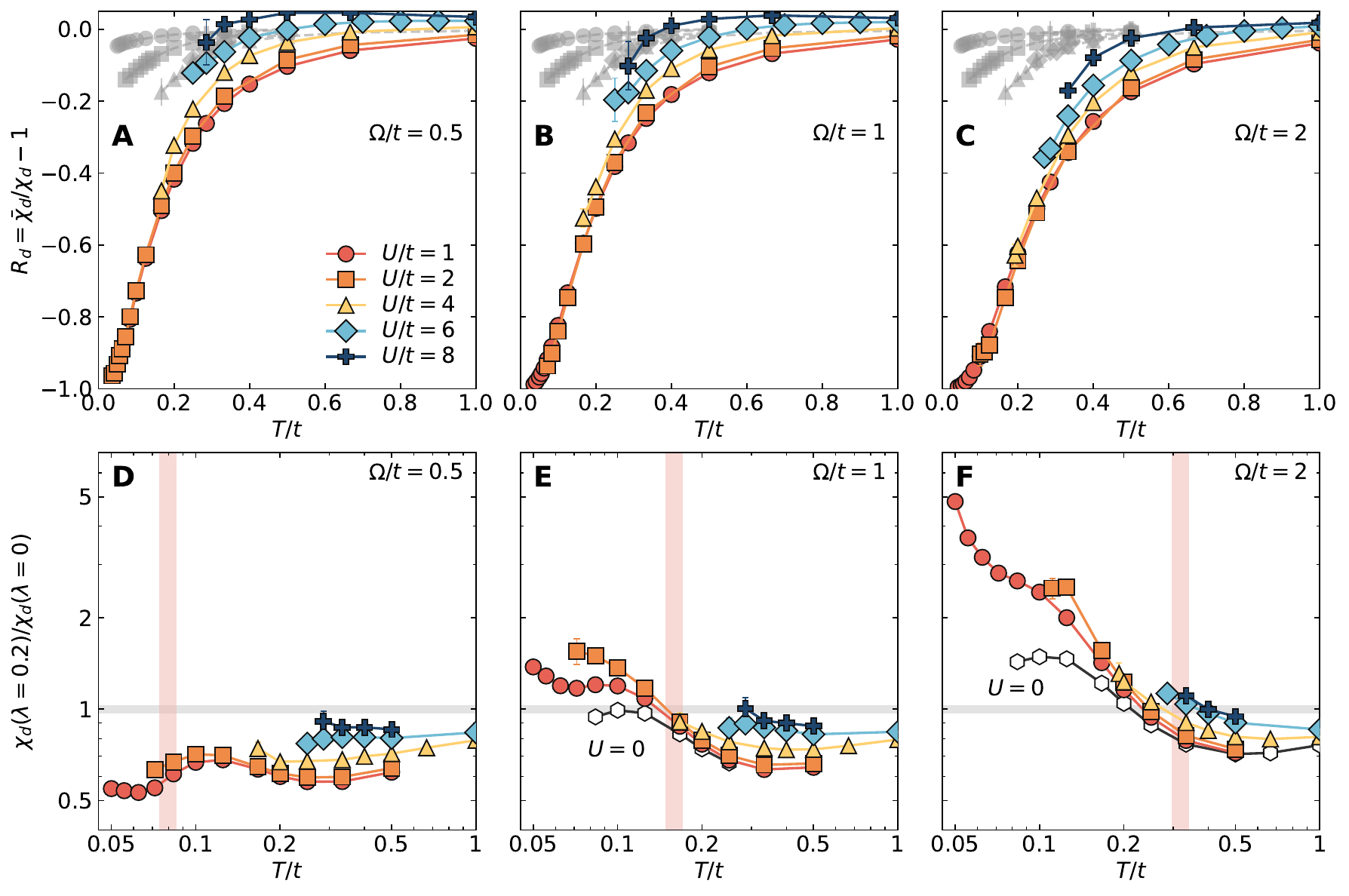}
    \caption{The same quantities as in Fig.~\ref{fig:main_vertex}A (upper row) and B (lower row) of the main text, but for $\Omega/t=0.5$ (A,D), 1 (B,E), and 2 (C,F). In A--C, colored solid curves have $\lambda=0.2$, and gray dashed curves have $\lambda=0$; symbols distinguish $U$. The gray horizontal bands mark unity, and the pink bands mark $T=\Omega/(2\pi)$. Other parameters are the same as in Fig.~\ref{fig:main_vertex}.}
    \label{fig:supp_vertex_u}
\end{figure}

Supplementary Fig.~\ref{fig:supp_vertex_u}A--C extends the $U$ dependence of $R_d\equiv\bar\chi_d/\chi_d-1$ to $\Omega/t=0.5,1,2$.
At high temperatures, $R_d$ is less negative at larger $U$, but it decreases upon cooling across the interaction range studied.
For $\Omega/t=1$ and $2$, the $U/t=2$ curve crosses below that at $U/t=1$ at low temperatures, consistent with the trend at $\Omega/t=4$ in Fig.~\ref{fig:main_vertex}A.
This low-temperature evolution is harder to resolve at smaller $\Omega$, where the regime $T\lesssim\Omega/(2\pi)$ is more difficult to access.

The phonon enhancement $\chi_d(\lambda=0.2)/\chi_d(\lambda=0)$ is also harder to resolve at smaller $\Omega$, as shown in Supplementary Fig.~\ref{fig:supp_vertex_u}D--F: temperatures must fall below $T\sim\Omega/(2\pi)$, marked by the pink bands, as discussed for Fig.~\ref{fig:main_omega_crossover} of the main text.
Smaller $\Omega$ also produces stronger finite-size effects, especially at weak $U$, as shown in Supplementary Fig.~\ref{fig:supp_size_ratio}.
Nevertheless, the overall $U$ dependence of the enhancement in panels D--F is consistent with that in Fig.~\ref{fig:main_vertex}B.

Inspired by the inverse-susceptibility structure of the particle--particle Bethe--Salpeter equation~\cite{PhysRevB.47.14599}, we define a scalar effective $d$-wave pairing interaction from the midpoint responses,
\begin{equation}
V_d(U,\lambda)=\bar\chi_d(U,\lambda)^{-1}-\chi_d(U,\lambda)^{-1}.
\label{eq:effective_pairing_interaction}
\end{equation}
\begin{samepage}
This scalar measure does not represent the full momentum- and frequency-dependent irreducible vertex, but we use it to characterize an effective $d$-wave pairing interaction: positive $V_d$ denotes attraction that enhances pairing relative to the dressed single-particle bubble~\cite{PhysRevB.39.839}.
We isolate the non-additive contribution through
\begin{equation}
S_d(U,\lambda)=V_d(U,\lambda)-V_d(0,\lambda)-V_d(U,0)+V_d(0,0).
\label{eq:pairing_nonadditivity}
\end{equation}
\end{samepage}
Here $S_d>0$ indicates that the combined effective $d$-wave attraction exceeds the sum of the separate phonon and Hubbard contributions, implying non-additivity.
Although $S_d$ can be negative at intermediate temperatures, we find positive $S_d$ at low temperatures for both $\Omega/t=2$ and $4$, as shown in Supplementary Fig.~\ref{fig:supp_nonadditivity}.
The non-additive contribution grows upon cooling and generally increases with $U$ at low temperatures, showing that the phonon and Hubbard repulsion cooperate in enhancing the effective $d$-wave attraction beyond their separate contributions.

\begin{figure}[tbp]
    \centering
    \includegraphics[width=\linewidth]{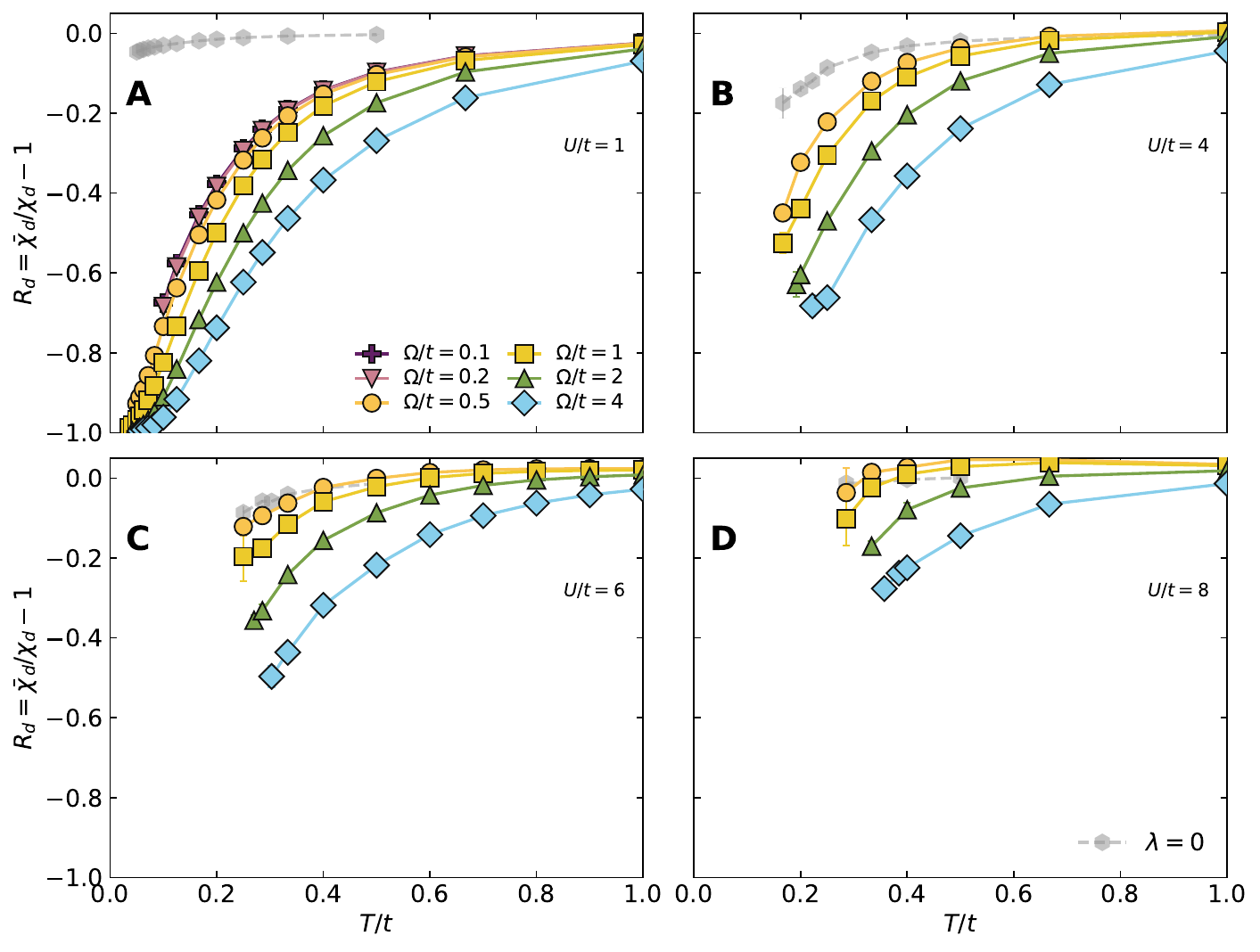}
    \caption{Temperature dependence of $R_d=\bar\chi_d/\chi_d-1$ at $U/t=1$ (A), 4 (B), 6 (C), and 8 (D). Colored solid curves distinguish different $\Omega$ at $\lambda=0.2$; gray dashed curves show $\lambda=0$. Parameters: $t'/t=-0.25$, $\langle n\rangle=0.8$; lattice size is $8\times8$.}
    \label{fig:supp_vertex_omega}
\end{figure}

Supplementary Fig.~\ref{fig:supp_vertex_omega} extends the frequency dependence of $R_d$ to $U/t=1,4,6,8$. At fixed temperature, larger $\Omega$ generally produces stronger $d$-wave vertex enhancement, as in Fig.~\ref{fig:main_omega_crossover}A of the main text.

The crossover at $U/t=2,4,6$ in Fig.~\ref{fig:main_omega_crossover}B--D extends to $U/t=1,8$ in Supplementary Fig.~\ref{fig:supp_omega_crossover}. The observed crossings also occur at temperatures of order $\Omega/(2\pi)$ across these interaction strengths, as in Fig.~\ref{fig:main_omega_crossover}B--D of the main text.
Small-$\Omega$ downturns in Supplementary Fig.~\ref{fig:supp_omega_crossover}A and Fig.~\ref{fig:main_omega_crossover}B of the main text appear to deviate from the common retardation behavior. However, the ratios near the downturns depend strongly and non-monotonically on lattice size, as shown in Supplementary Fig.~\ref{fig:supp_size_ratio}A,C. These downturns therefore do not establish a bulk trend. At the relatively large phonon frequency $\Omega/t=4$, all three sizes instead show strong enhancement upon cooling below $2\pi T/\Omega\sim1$, despite quantitative differences, as shown in Supplementary Fig.~\ref{fig:supp_size_ratio}B,D.

\begin{figure}[tbp]
    \centering
    \includegraphics[width=\linewidth]{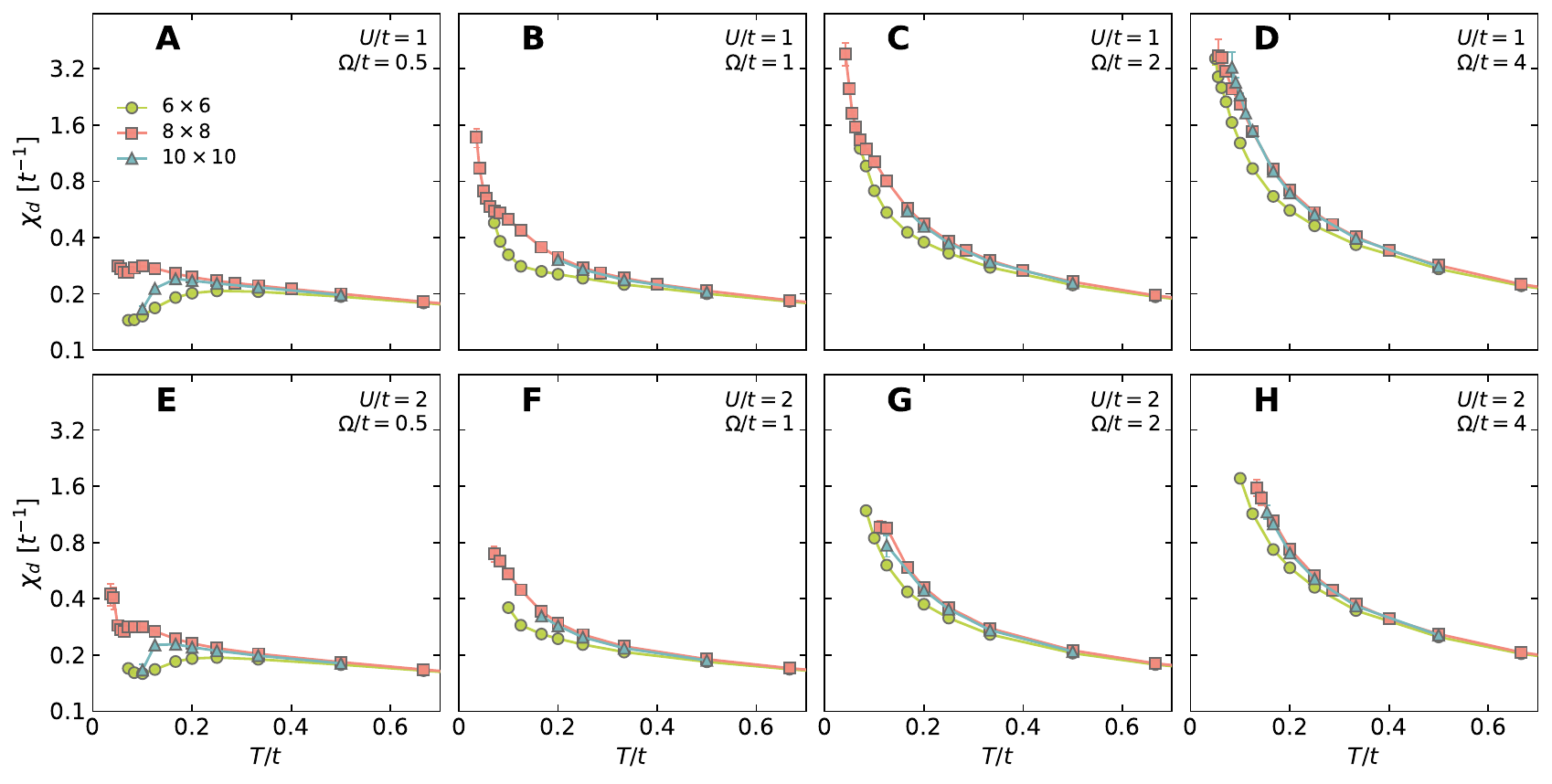}
    \caption{Temperature dependence of the $d$-wave midpoint correlator, $\chi_d$, at different lattice sizes. (A--D) $U/t=1$ and (E--H) $U/t=2$, each ordered by $\Omega/t=0.5,1,2,4$. Colors denote lattice size. Parameters: $t'/t=-0.25$, $\lambda=0.2$, $\langle n\rangle=0.8$.}
    \label{fig:supp_size_omega}
\end{figure}

{\widowpenalty=10000
Supplementary Fig.~\ref{fig:supp_size_omega} compares $\chi_d$ on $6\times6$, $8\times8$, and $10\times10$ lattices. At low temperature, the $6\times6$ results lie below those for larger lattices, but the $8\times8$ and $10\times10$ results closely agree at large $\Omega$. At smaller $\Omega$, discrete single-particle level spacing could become more relevant~\cite{Bormann1991FiniteSize}. Finite-size effects make it difficult to resolve substantial phonon enhancement of pairing, since the temperatures at which the $8\times8$ and $10\times10$ responses closely agree are still too high, as discussed above.\par}

\begin{figure}[tbp]
    \centering
    \includegraphics[height=0.70\textheight,keepaspectratio]{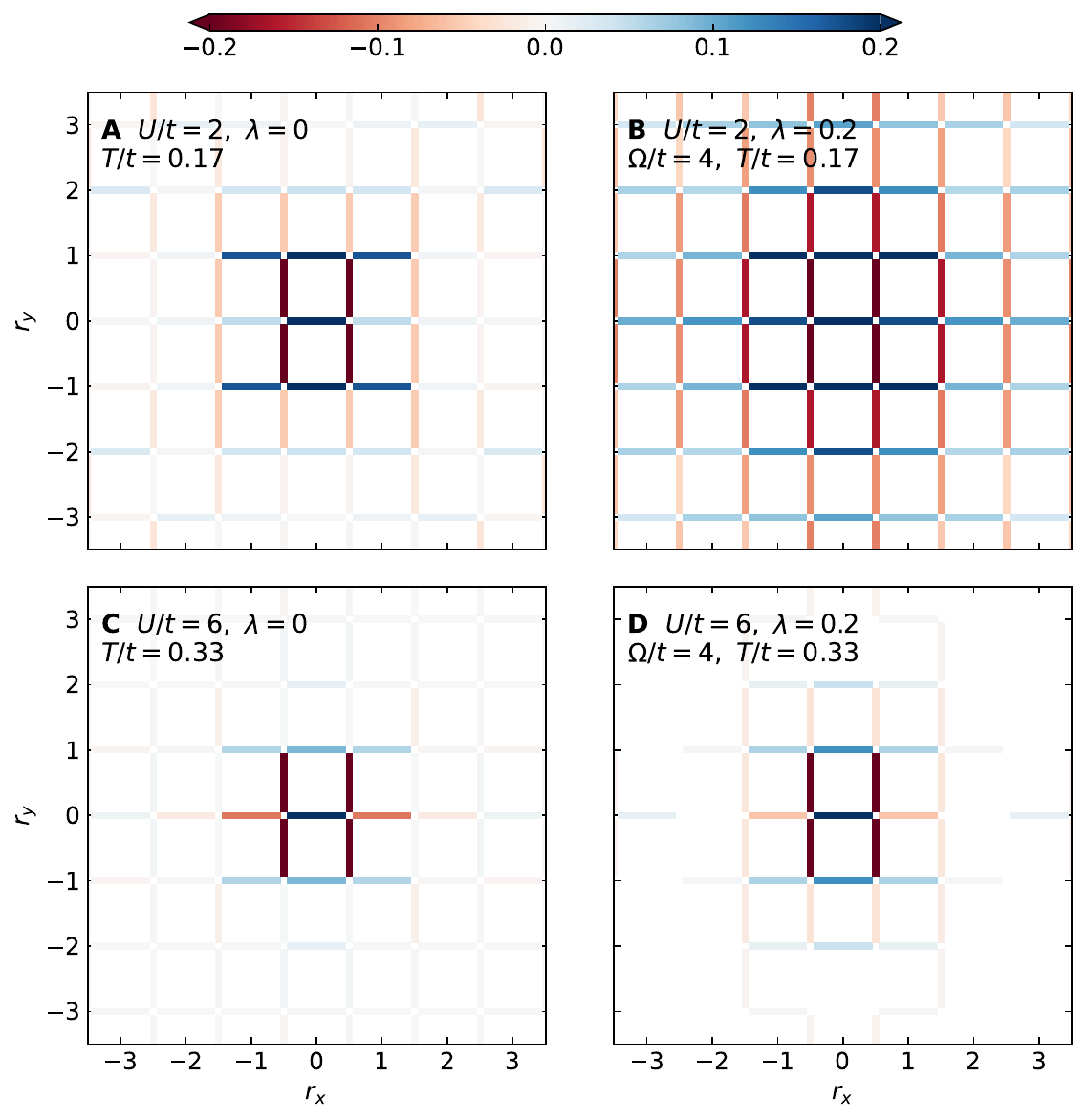}
    \caption{Real-space $d$-wave bond-pair correlations, $S_{\alpha,\hat x}(\mathbf r,\beta/2)/S_{\hat x,\hat x}(\mathbf 0,\beta/2)$, where $S_{\alpha,\alpha'}(\mathbf r,\tau)=N_s^{-1}\sum_i\langle\mathcal T_\tau\Delta_\alpha(\tau,\mathbf r_i+\mathbf r)\Delta_{\alpha'}^\dagger(0,\mathbf r_i)\rangle$, the central $\hat x$ bond is the reference, and $\alpha=\hat x,\hat y$. (A,B) $U/t=2$ at $T/t=1/6$; (C,D) $U/t=6$ at $T/t=1/3$. The left column has $\lambda=0$; the right column has $\lambda=0.2$ and $\Omega/t=4$. Parameters: $t'/t=-0.25$, $\langle n\rangle=0.8$; lattice size is $8\times8$. Bonds are shown when the magnitude of the correlation exceeds twice its error bar. The color scale is clipped at $\pm0.2$, so the reference bond is saturated in every panel.}
    \label{fig:supp_realspace}
\end{figure}

Supplementary Fig.~\ref{fig:supp_realspace} displays the spatial structure of the bond-pair correlations. At $U/t=2$, the $B_{1g}$ phonon turns short-ranged correlations into a $d$-wave pattern spanning the lattice. At $U/t=6$, although the accessible temperature is much higher, the phonon increases their range and makes the $d$-wave pattern more clearly defined.

\begin{figure}[tbp]
    \centering
    \includegraphics[width=\linewidth]{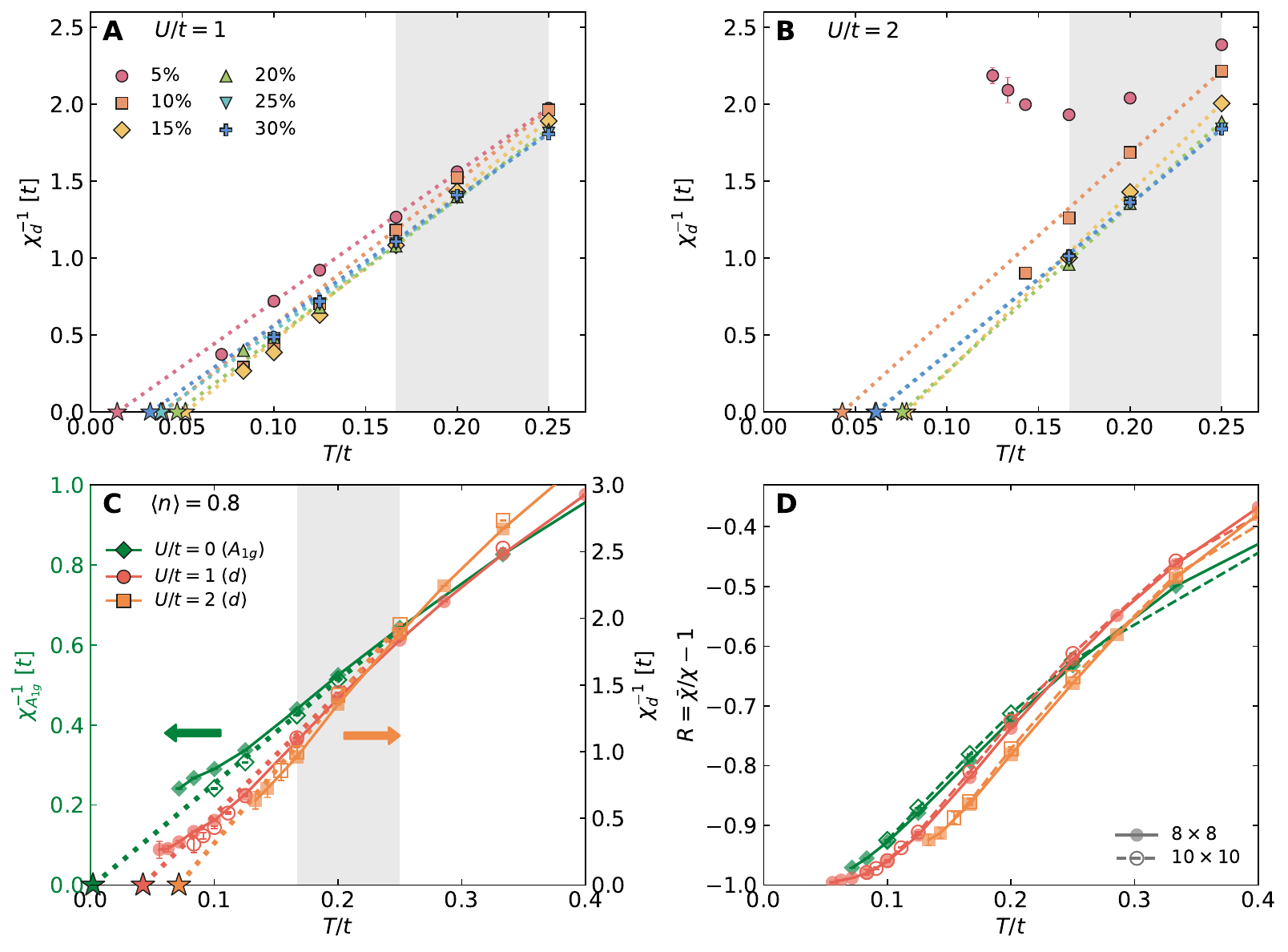}
    \caption{Curie--Weiss fits behind Fig.~\ref{fig:main_doping}. (A,B) $\chi_d^{-1}$ versus temperature at dopings $1-\langle n\rangle=5$--$30\%$ for $U/t=1$ (A) and $2$ (B) on $8\times8$ lattices. Colors and symbols denote doping (see legend). Dotted lines are the Curie--Weiss fits at $\beta t=4,5,6$ (gray band), and stars mark their zeros $T_0$, shown in Fig.~\ref{fig:main_doping}A. (C) Inverse pairing susceptibilities at $\langle n\rangle=0.8$: $\chi_{A_{1g}}^{-1}$ at $U=0$ (left axis; Eq.~\eqref{eq:a1g_leading}) and $\chi_d^{-1}$ at $U/t=1,2$ (right axis); arrows point each set of curves to its axis. Dotted lines are Curie--Weiss fits to the $10\times10$ data at $\beta t=4,5,6$ (gray band), and stars mark their zeros $T_0$, shown in Fig.~\ref{fig:main_doping}B. (D) Vertex ratio $R=\bar\chi/\chi-1$: $R_d$ at $U/t=1,2$ and $R$ of the most strongly enhanced $A_{1g}$ combination at $U=0$. In (C,D), semi-transparent filled symbols denote $8\times8$ and open symbols $10\times10$; lines connect the $8\times8$ data (solid) and, in (D), the $10\times10$ data (dashed). Parameters: $t'/t=-0.25$, $\lambda=0.2$, $\Omega/t=4$.}
    \label{fig:supp_u_tc}
\end{figure}

Supplementary Fig.~\ref{fig:supp_u_tc} collects the Curie--Weiss fits behind Fig.~\ref{fig:main_doping}. Panels A and B show $\chi_d^{-1}$ at every doping for $U/t=1$ and $2$, with the $8\times8$ fits that give Fig.~\ref{fig:main_doping}A. Panels C and D compare the phonon-only model ($U=0$, $\lambda=0.2$) with $U/t=1,2$ at $\langle n\rangle=0.8$. At $U=0$ the leading pairing channel is $A_{1g}$ rather than $d$-wave (Supplementary Fig.~\ref{fig:supp_a1g}). Because $\Delta_s$ and $\Delta_{f_d^2}$ belong to the same representation and mix, we consider the general $A_{1g}$ pair field $\Delta_{\mathbf v}=v_s\Delta_s+v_{f_d^2}\Delta_{f_d^2}$ with $\mathbf v^{\mathsf T}\mathbf v=1$ and the $2\times2$ midpoint susceptibility matrix
\begin{equation}
\chi_{\alpha\alpha'}=\frac{\beta}{N_s}\langle\mathcal T_\tau\Delta_\alpha(\beta/2)\Delta_{\alpha'}^\dagger(0)\rangle,\qquad\alpha,\alpha'\in\{s,f_d^2\},
\label{eq:a1g_matrix}
\end{equation}
whose diagonal elements are $\chi_s$ and $\chi_{f_d^2}$ of Eq.~\eqref{eq:pair_midpoint}. The most strongly fluctuating combination maximizes $\mathbf v^{\mathsf T}\chi\mathbf v$, so its susceptibility is the largest eigenvalue
\begin{equation}
\chi_{A_{1g}}=\tfrac12\Bigl[\chi_s+\chi_{f_d^2}+\sqrt{(\chi_s-\chi_{f_d^2})^2+4\chi_{s f_d^2}^2}\,\Bigr].
\label{eq:a1g_leading}
\end{equation}
Within each resample the measured cross term is held to the Cauchy--Schwarz bound $|\chi_{s f_d^2}|\le(\chi_s\chi_{f_d^2})^{1/2}$. Because the operator normalizations differ between channels, the absolute values of $\chi_{A_{1g}}$ and $\chi_d$ are not comparable; we compare their temperature dependence and their Curie--Weiss temperatures. The temperature dependence in Supplementary Fig.~\ref{fig:supp_u_tc}C points to a lower transition scale for the leading $A_{1g}$ channel at $U=0$ than for the $d$-wave channel at $U/t=1,2$.

For the vertex ratio, we form the bare matrix $\bar\chi_{\alpha\alpha'}$ from the same Green's functions as Eq.~\eqref{eq:pair_bare_bubble}, with the form factors $1$ and $f_d(\mathbf k)^2$, so that $\bar\chi_{s f_d^2}=\bar\chi_d$. The combination $\Delta_{\mathbf u}=u_s\Delta_s+u_{f_d^2}\Delta_{f_d^2}$ with the strongest vertex enhancement maximizes $\rho=\mathbf u^{\mathsf T}\chi\mathbf u/\mathbf u^{\mathsf T}\bar\chi\mathbf u$; its $\rho_{\max}$ is the largest root of $\chi\mathbf u=\rho\,\bar\chi\mathbf u$, and $R=1/\rho_{\max}-1$ reduces to $R_X$ of Eq.~\eqref{eq:relative_vertex_enhancement} for a single channel. Supplementary Fig.~\ref{fig:supp_u_tc}D likewise shows $R_d$ at $U/t=1,2$ tending toward $-1$ at higher temperatures than the $A_{1g}$ vertex ratio at $U=0$.

To relate the two combinations, we write $\chi=\chi_{A_{1g}}\mathbf v\mathbf v^{\mathsf T}+\chi_2\mathbf v_2\mathbf v_2^{\mathsf T}$, where $\mathbf v_2$ is the second unit eigenvector and $\chi_{A_{1g}}\ge\chi_2\ge0$. The Cauchy--Schwarz inequality gives $(\mathbf w^{\mathsf T}\mathbf u)^2\le(\mathbf w^{\mathsf T}\bar\chi^{-1}\mathbf w)(\mathbf u^{\mathsf T}\bar\chi\mathbf u)$, with equality when $\mathbf u=\bar\chi^{-1}\mathbf w$. Applying this inequality for $\mathbf w=\mathbf v$ and $\mathbf v_2$ yields
\begin{equation}
\chi_{A_{1g}}\,\mathbf v^{\mathsf T}\bar\chi^{-1}\mathbf v\le\rho_{\max}\le\chi_{A_{1g}}\,\mathbf v^{\mathsf T}\bar\chi^{-1}\mathbf v+\chi_2\,\mathbf v_2^{\mathsf T}\bar\chi^{-1}\mathbf v_2,
\label{eq:rho_bounds}
\end{equation}
where the lower bound is $\rho$ at $\mathbf u=\bar\chi^{-1}\mathbf v$ without its $\chi_2$ term. If only the leading mode diverges at the transition, $\chi_{A_{1g}}\to\infty$ while $\chi_2$ and the positive-definite $\bar\chi$ stay finite. Both bounds then grow as $\chi_{A_{1g}}$, so $\chi_{A_{1g}}$ and $\rho_{\max}$ diverge at the same temperature, and the maximizing combination becomes $\mathbf u=\bar\chi^{-1}\mathbf v$.

\begin{figure}[tbp]
    \centering
    \includegraphics[width=\linewidth]{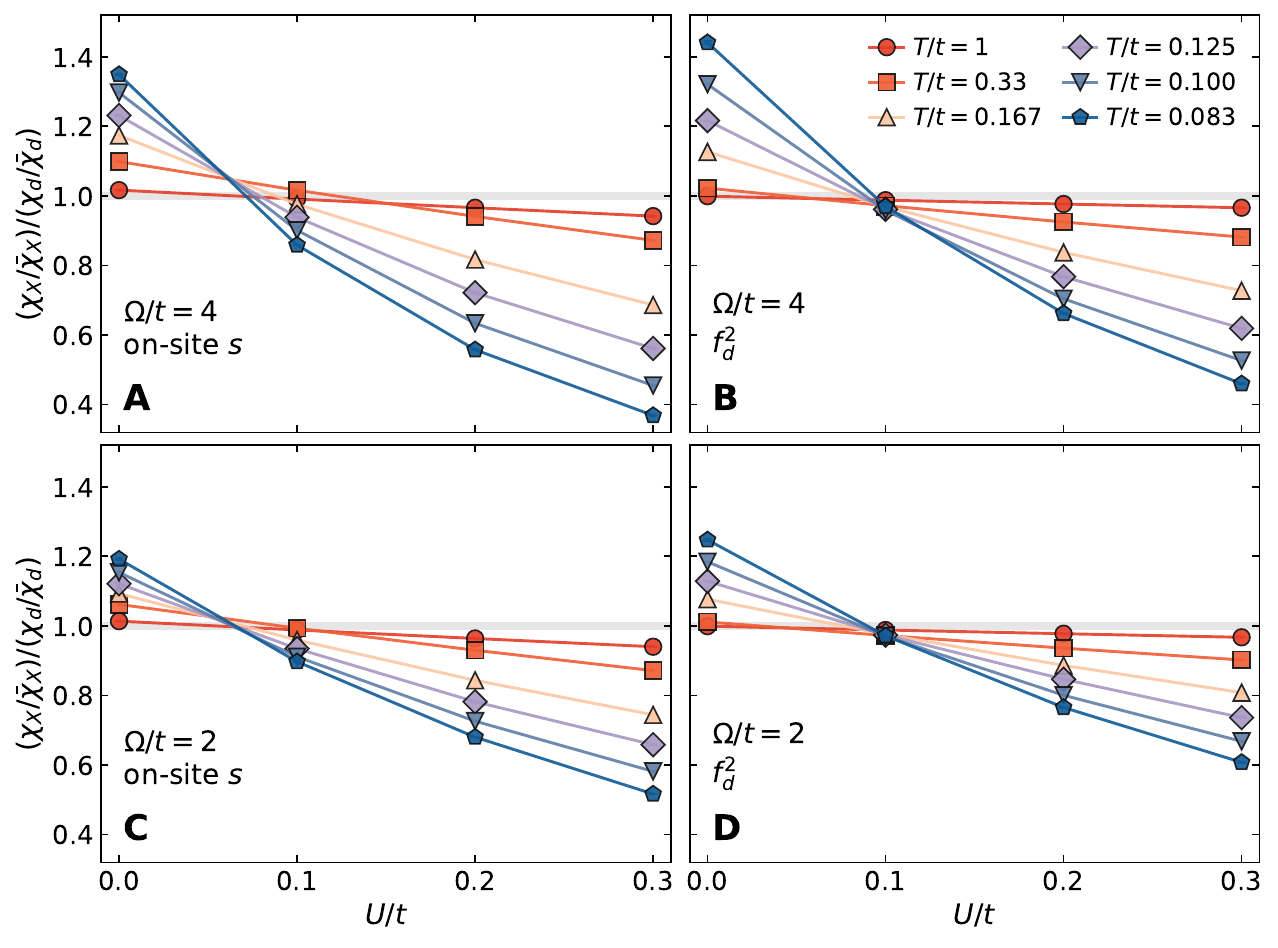}
    \caption{Ratio of the pairing vertex enhancements, $(\chi_X/\bar\chi_X)/(\chi_d/\bar\chi_d)$, versus $U/t$ for the on-site $s$-wave channel (A,C) and the $f_d^2$-weighted $A_{1g}$ channel (B,D). (A,B) $\Omega/t=4$; (C,D) $\Omega/t=2$. Unity denotes equal enhancement. Parameters: $t'/t=-0.25$, $\lambda=0.2$, $\langle n\rangle=0.8$; lattice size is $8\times8$.}
    \label{fig:supp_uc_vertex}
\end{figure}

\begin{figure}[tbp]
    \centering
    \includegraphics[width=\linewidth]{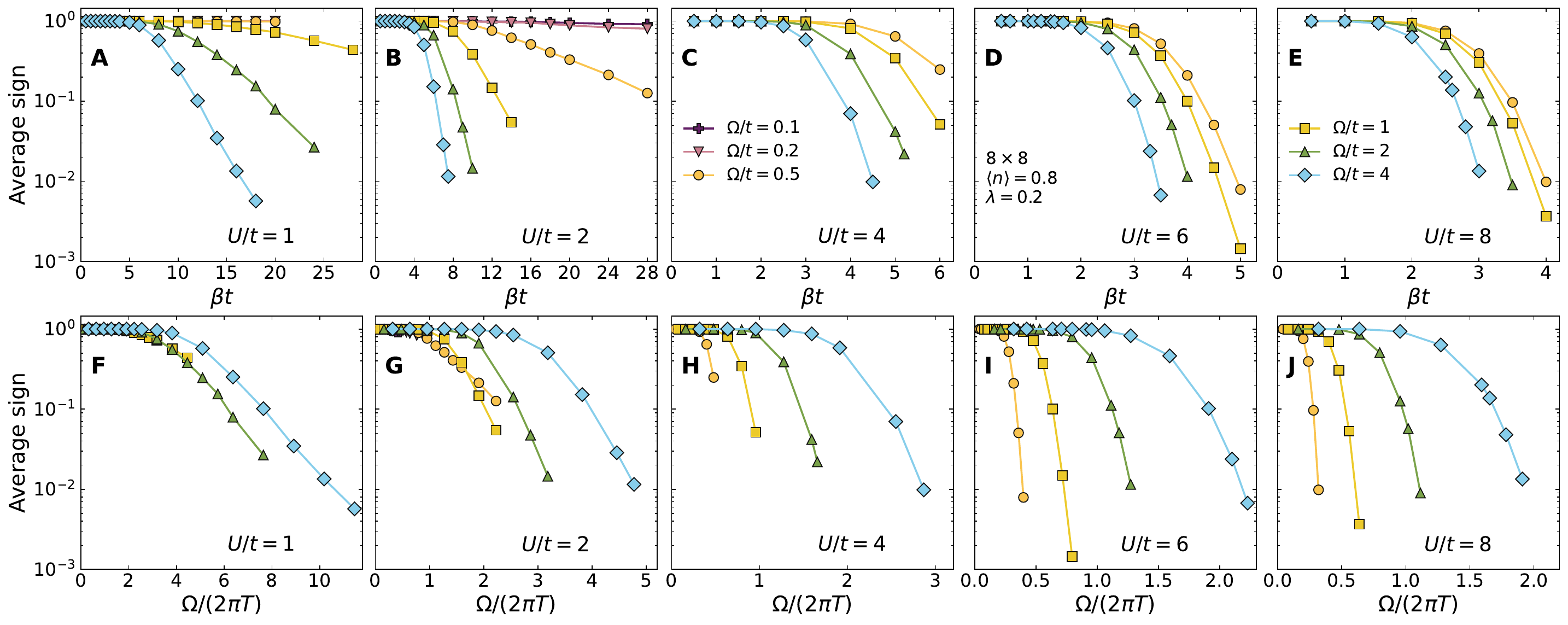}
    \caption{Average fermion sign at $\langle n\rangle=0.8$ on an $8\times8$ lattice, plotted versus $\beta t$ in (A--E) and $\Omega/(2\pi T)$ in (F--J). Columns correspond to $U/t=1,2,4,6,8$. Parameters: $t'/t=-0.25$, $\lambda=0.2$.}
    \label{fig:supp_sign_temperature}
\end{figure}

\begin{figure}[tbp]
    \centering
    \includegraphics[width=\linewidth]{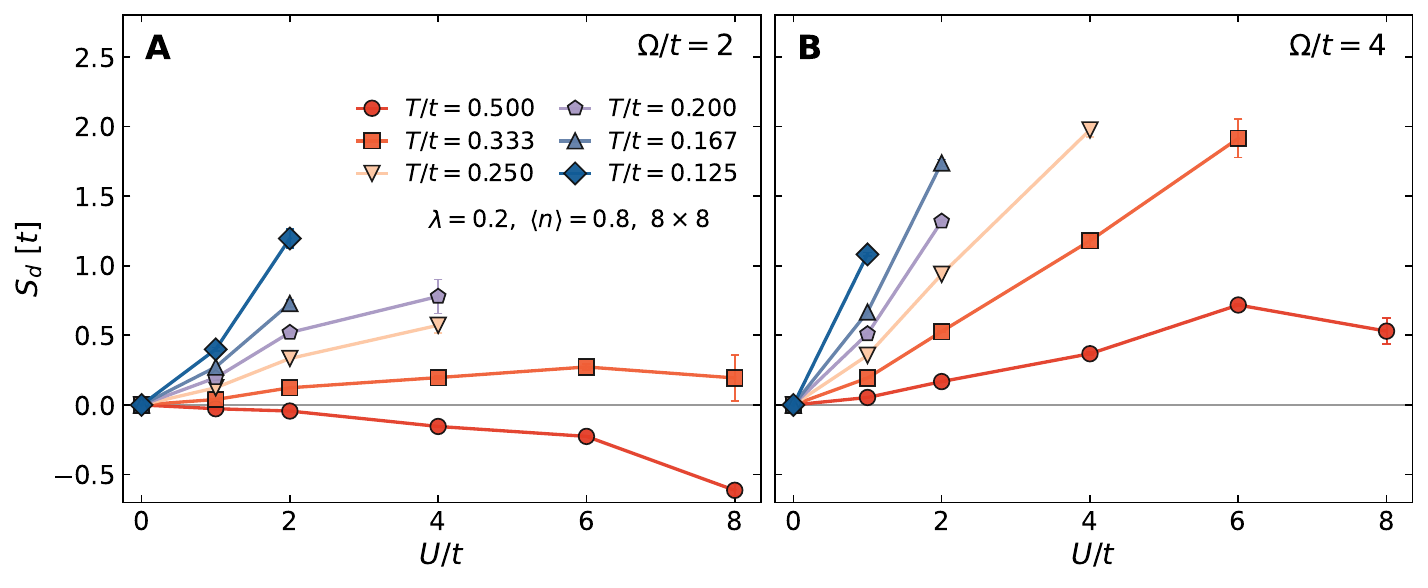}
    \caption{Non-additivity of the effective $d$-wave pairing interaction, $S_d=V_d(U,\lambda)-V_d(0,\lambda)-V_d(U,0)+V_d(0,0)$, with $V_d=\bar\chi_d^{-1}-\chi_d^{-1}$, for $\Omega/t=2$ (A) and 4 (B). Parameters: $\lambda=0.2$, $t'/t=-0.25$, $\langle n\rangle=0.8$; lattice size is $8\times8$.}
    \label{fig:supp_nonadditivity}
\end{figure}

\begin{figure}[tbp]
    \centering
    \includegraphics[width=\linewidth]{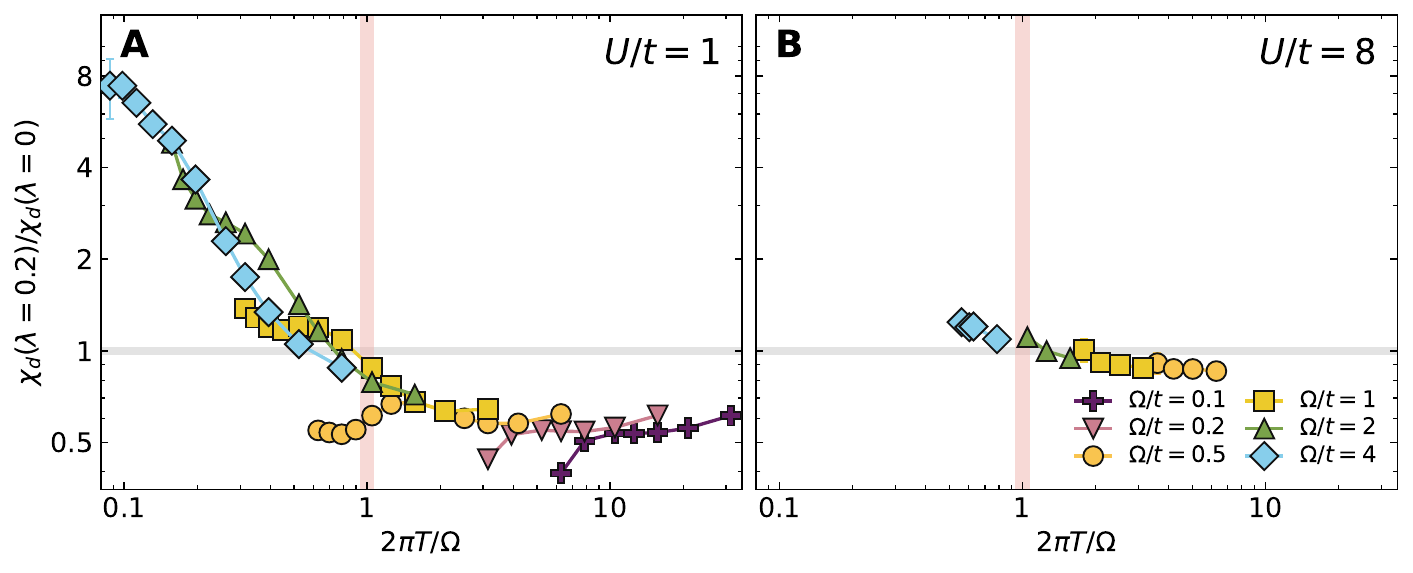}
    \caption{The same quantity as in Fig.~\ref{fig:main_omega_crossover}B--D of the main text, but for $U/t=1$ (A) and 8 (B).}
    \label{fig:supp_omega_crossover}
\end{figure}

\begin{figure}[tbp]
    \centering
    \includegraphics[width=\linewidth]{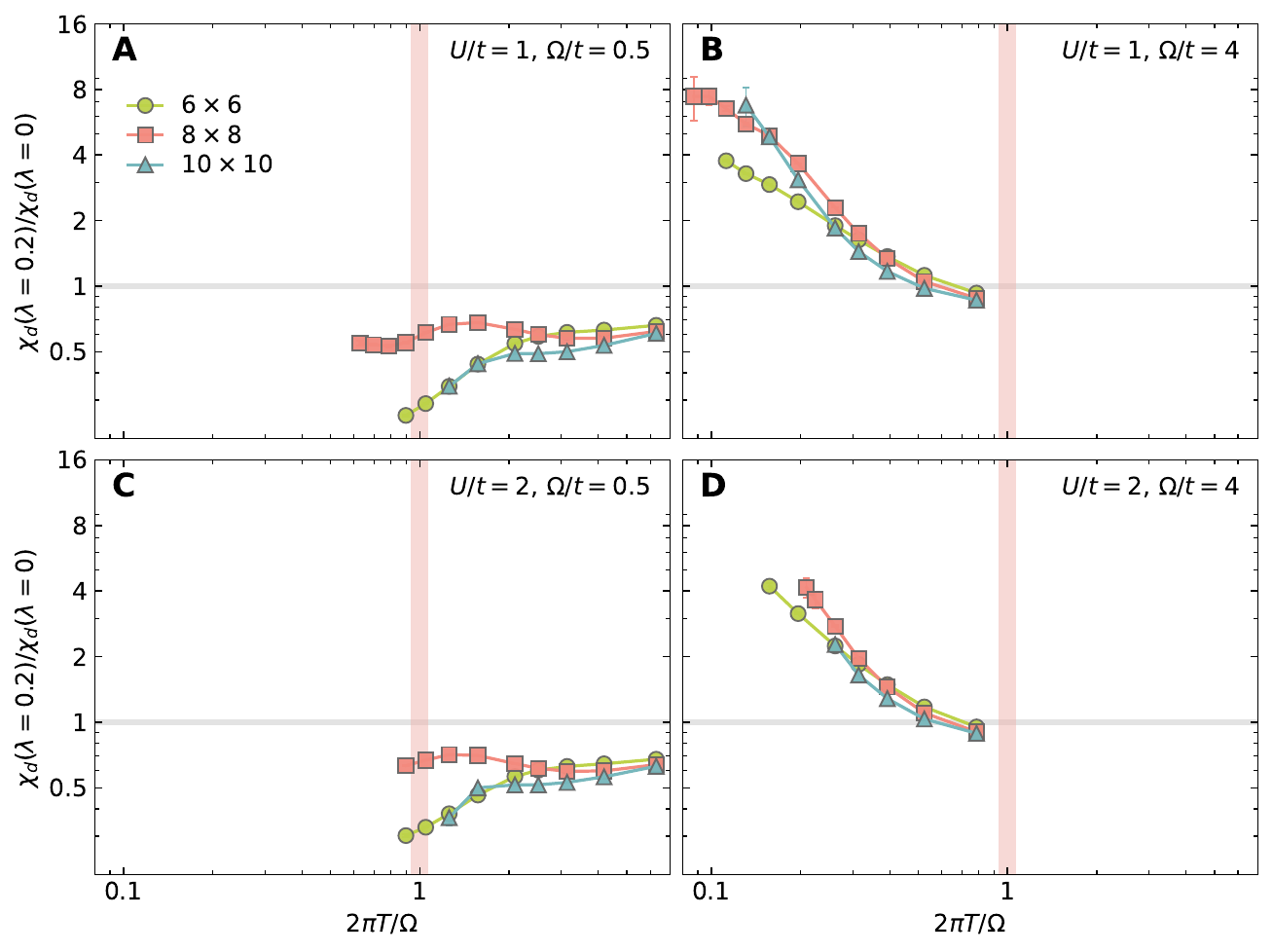}
    \caption{Size dependence of the $d$-wave midpoint-correlator ratio, $\chi_d(\lambda=0.2)/\chi_d(\lambda=0)$, versus $2\pi T/\Omega$, with a $\lambda=0$ Hubbard reference at the same lattice size. (A,B) $U/t=1$; (C,D) $U/t=2$. (A,C) $\Omega/t=0.5$; (B,D) $\Omega/t=4$. The gray horizontal band marks unity; the pink vertical band marks $T=\Omega/(2\pi)$. Parameters: $t'/t=-0.25$, $\langle n\rangle=0.8$.}
    \label{fig:supp_size_ratio}
\end{figure}

\end{document}